# Complex Observation in Electron Microscopy VIII: Novel Hilbert Phase-plates to Maximize Phase-contrast Sensitivity

Kuniaki Nagayama

N-EM Laboratories Inc.
505 CIC Tamachi Campus, 3-3-6 Shibaura, Minato, Tokyo, 108-0023, Japan
E-mail: nagayama@nips.ac.jp

Abstract

  Phase-plate transmission electron microscopy has recently regressed since a report that Volta phase-plate phase-contrast is less sensitive than non-phase-plate phase contrast, which leads to conventional defocusing phase-contrast. What about Hilbert phase-plate phase-contrast?  We report that the Hilbert phase-plate method can survive if two experiments using a pair of symmetric Hilbert phase-plates, of which phase is set to a value smaller than π, are combined.  Three phase-contrast methods using the symmetric Hilbert phase-plates and Zernike phase-plate representing Volta phase-plate and Scherzer defocus respectively were compared in sensitivity theoretically relying on a contrast transfer theory and computationally on a simulator specifically designed for phase plate transmission electron microscopy. For the two phase-plate phase-contrasts, the phase that gives the highest sensitivity was searched for by changing the phase-plate phase. As a result, the symmetric Hilbert phase-plates phase-contrast was found to outperform Scherzer defocus phase-contrast in the phase around π/2. On the other hand, Zernike phase-plate phase-contrast was found considerably inferior to Scherzer defocus phase-contrast in the entire phase range from 0 to π. Furthermore, the novel Hilbert phase-plate method was compared with complex observation transmission electron microscopy, which also requires two experiments, and the origin of the higher sensitivity of symmetric Hilbert phase-plates phase-contrast was examined.

# 1. Introduction

Two contrasting phase-contrast (PC) methods are used in transmission electron microscopy (TEM): one uses the lens defocusing to generate phase-contrast (PC) [1] and the other utilizes the phase shift induced by a phase-plate (PP) placed at the back focal plane (BFP) [2]. Among the various PPs that have been proposed so far [3], two representative PPs; ZPP generating Zernike PC (ZPC) *[1] [4] and HPP generating Hilbert-differential phase-contrast*[2](HDPC) [5], are focused in this study together with the traditional method using Scherzer defocus [6]. The term "contrast", in this paper, refers to the deviation from the background bright field (normalized as *1*) in the vision.

While the experimental difficulties with PPs have mainly been discussed in terms of image disturbance due to electric charging [7], a recent paper has focused on another difficulty, namely sensitivity reduction, where the discrepancy in sensitivity with and without a Volta PP (VPP) was investigated in a 300 kV biological TEM with a 5 $nm^{-1}$ information limit [8]. As a result, a 10% signal loss in the low-frequency range (around 1 $nm^{-1}$) and a 60% signal loss in the high-frequency range (around 5 $nm^{-1}$) were found even with a 3 nm-thick ultrathin PP. As the PP was ultra-thin, the signal loss was attributed to a type of detective quantum efficiency (DQE) supposing to be derived from positional fluctuations of electron induced surface charge [8] rather than the electron scattering in the PP [4]. The VPP used was a form of Zernike-type PPs originally proposed as a special measure to avoid charge-induced image disturbance in PPs [9, 10], but the newly identified difficulty of DQE-type must be even more serious than the image disturbance, as it can be considered an unavoidable flaw of PPs in general. In retrospect, it seems to highlight the sensitivity advantage of the Scherzer defocus method [6], which does not rely on PPs to recover PCs.

Phase plates studied in this report belong to two types; one is ZPP-type, which covers the aperture entirely and the other HPP-type, which, on the other hand, half covers the aperture. As one of the inventors of HPP, the author became interested in how the DQE loss in question might be in the HPP-type. In the course of surveys on the subject, he found two papers of interest [11,12].

The two papers have both concerned an extension of HPP by focusing on how PC signals are altered when the HPP phase is changed from π to smaller values [11,12]. They conveyed that deviation of the phase from π inevitably results in a mixing of two type of PCs, normal-type represented by ZPC and differential-type represented by HDPC, but rather a positive aspect arising from it. The Koeck paper [11] reported that, for the phase *φ* = π/2, the mixing of two types of PCs could be eliminated by a complex-number inverse filtering in spectral space designated by ***k*** (the term spectral space is preferred in this study over reciprocal space or frequency space being aware of the Fourier pair, real space designated by ***r***) and can recover normal-type PC with a spatial resolution higher than that for *φ* = π. The Edgecombe paper [12] treated the PP phase and the distance of the optical axis from the half-planar HPP edge as variables and examined to what extent the reproduction of object images was faithful, and found that, as in the Koeck paper, *φ* = π/2 was favored in the reproduction of copy images of the specimen.

This study shares with the preceding two papers on turning the troublesome PC mixing into a positive aspect but rather focuses onto the sensitivity improvement instead of the resolution improvement or image-copy reproduction. Our way for avoiding the pitfall of the PC mixing, which occurs with the HPP using a PP phase $\varphi$ smaller than $\pi$ (termed as $\varphi$-HPP), is to utilize two TEM experiments, as like the two-experiment scheme used in complex observation (COBS) [13, 14].

PC TEM translates phase fluctuations in the object into amplitude modulations in the image. The phase-contrast transfer function (PCTF or short CTF) illustrates who well the PC TEM does this [15] . Using the CTF, which is a kind of filter in spectral space and useful for searching which frequencies are transferred to the final image through the lens system used, the performance of a novel PC scheme using a symmetric pair of $\varphi$-HPPs is reported in this paper from the angle of sensitivity.

The rest of the paper is organized as follows.

In Sect. 2, we describe how the variation of the PP phase $\varphi$ affects imaging properties in TEM with $\varphi$-HPP. For comparison, $\varphi$-ZPP is also investigated. Phase-contrast and sensitivity in imaging are defined and a theoretical formulation of the sensitivity for $\varphi$-HPP is presented, first under an ideal condition without electron dose losses due to the PP and next under realistic conditions with electron dose losses. Here, as an extension of a single $\varphi$-HPP, a pair of symmetric $\varphi$-HPPs, which expresses the freedom of whether the half-planar PP comes on the left side or on the right side of the optical axis, is introduced and a highly sensitive two-experiment scheme using a pair of symmetric $\varphi$-HPPs (termed as twin $\varphi$-HPPs) is proposed.

In Sect. 3, CTFs for three PCs are developed by including actual instrument parameters such as wave aberrations of objective lens and hardware instabilities of TEM equipment, both of which severely impair the TEM performance. Here, a phenomenon of mixing of two types of CTFs, one typical in defocusing and the other in ZPP, is reported for $\varphi$-HPP PC. It is finally led to the high performance of the novel PP method using twin $\varphi$-HPPs.

In Sect. 4, in order to intuitively understand the imaging performance, CTFs corresponding to the three PC methods described in the previous section are visualized. One-dimensional (1D) functions of a product of three functions (CTF proper and two envelopes) in a radial direction are usually illustrated for CTF curves. However, as they are essentially two-dimensional (2D) functions defined in the spectral space with frequency vector ***k***, 2D integration of the three-function product that corresponds to sensitivity in this paper finally becomes a 1D integration of a four-function product with an additional product of $k$. The numerical results of the sensitivity integral are then illustrated as sensitivity ratios among the three PC methods.

In Sect. 5, theoretical evaluations for the sensitivity are practically verified with a TEM simulator for a specific specimen, a single atom of gold and the high sensitivity of twin $\varphi$-HPPs, when $\varphi$ is around $\pi/2$, is confirmed in comparison with the other two PC methods, the Scherzer defocus and the $\varphi$-ZPP.

In Sect. 6, corresponding to Sect. 5, the experimental algorithm with two-experiment scheme with twin $\varphi$-HPPs is illustrated with gold atom simulated images. The algorithm is visualized as a step-by-step flow of gold atom images, and finally the high-sensitivity image obtained as a unified one is compared with the other two PC images, the Scherzer defocus and $\pi/2$-ZPP image.

In Sect. 7, the pros and cons of first-order COBS realized with use of the symmetric twin $\varphi$-HPPs is discussed in comparison with traditional second-order COBS that uses an asymmetric combination of the prime-focus defocusing and $\pi/2$-ZPP PC, together with other subjects such as ways to avoid phase plate charging.

*1, Various variations for the ZPP method, which was proposed as a method to avoid the drawback of the defocusing method of low sensitivity in the low-frequency range, have been proposed [3]. The original ZPP [4] is a micro-device consisting of an amorphous carbon film with a micro-hole center mounted on an aperture diaphragm with an aperture of about 100 µm-diameter. The performance of the ZPP depends on the size of the micro-hole center. The smaller the mic-holes, the better the recovery in the low-frequency range and the higher the PC for large objects. However, the micro-holes cannot be smaller than, for example, 1 µm in order to prevent the PP charging. Hole-free PP (HFPP) [9] and Volta PP (VPP) [10] have been proposed to improve on this point and enabled recovery in the very low-frequency range. PC simulations using amorphous carbon spheres with diameters of 1 nm-20 nm [3b] have shown that among the PPs proposed so far, the HFPP has the best low-frequency range recovery and exhibits the highest PC for all amorphous carbon spheres used. With this in mind, in this report, the micro-hole of the ZPP was set to infinitesimal, namely an ideal ZPP was assumed, so to speak. The difference between apparent contrast, as used in Ref. 3b, and the sensitivity contrast used in this report, which is defined by considering a wide frequency range, will be discussed in Discussion.

*2, HDPC images resemble those obtained with the differential interference contrast (DIC), which is nowadays an indispensable tool in light microscopy [16], but actually are not the derivatives of images. The term "Hilbert-differential" is used to emphasize the pseudo-nature of differential-like images of HDPC. As a matter of fact, HDPC is nothing other than the Hilbert transform of the normal-type PC realized as defocus contrast or ZPC [5b].

## 2. Image Contrast with a $\varphi$-Hilbert Phase-plate

### 2.1 Optimum phase for $\varphi$-HPP under no electron dose loss condition

To make the framework of the contrast theory as clear as possible, we begin developing the theory under an ideal condition of no lens aberration and no electron dose loss due to the PP.

A $\pi$-HPP has a simple form as shown in Fig. 1a, where a PP with phase $\pi$ covers the left half-plane of the aperture at BFP. To avoid the direct beam (zero-order beam) of the un-scattered electrons colliding with the PP, the optical axis is set as close as to the PP but without touching it.

In this setting, the TEM yields a differential-like image [5] with a relief feature resembling DIC in light microscopy [16]. What will happen if the PP phase is set to any value φ smaller than π (Fig. 1b)? Though two recent papers partially answered this question [11, 12], we will answer the question from a different angle.

The φ-PP generally adds a uniform phase φ to the exit wave as,

$$exp(i\ \varphi) = cos\ \varphi + i\ sin\ \varphi, \qquad (1)$$

where, the PP is assumed to be a pure phase object with no electron dose loss.

The exit wave modulator $P(\mathbf{k})$ corresponding to φ-HPP shown in Fig. 1b, where a half-planar PP is set on the left side of the optical axis, is given by

$$P(\mathbf{k}) = A(\mathbf{k})\ [1 \cdot (1 + sgn(k_x))/2 + exp(i\ \varphi) \cdot (1 - sgn(k_x))/2]$$

$$= 1/2\ A(\mathbf{k})\ [\ (1 + cos\varphi) + (1 - cos\varphi)sgn(k_x) + i\ sin\varphi(1 - sgn(k_x))], \qquad (2)$$

where $A(\mathbf{k})$ corresponds to an aperture function, $(1 + sgn(k_x))/2$ to a left half-plane, $(1 - sgn(k_x))/2$ to a right half-plane, and $sgn(k_x)$ to a sign function ($sgn(k_x)=1$, $k_x \geq 1$; $sgn(k_x)=-1$, $k_x <1$), and · denotes an explicit form of multiplication. Note that our convention uses the absolute value for the vectorial frequency $\mathbf{k}$ (spectral space 2D coordinates), being expressed as $|\mathbf{k}| = k = 1/\lambda$, where $\lambda$ is the wavelength of electron wave determined by the acceleration voltage employed. This convention follows that for Fourier transforms (FTs) [17] as

$$\mathcal{F}[f(\mathbf{r})] = \int f(\mathbf{r}) e^{-i2\pi \mathbf{k}\mathbf{r}}\ d\mathbf{r} = F(\mathbf{k}), \qquad (3)$$

$$\mathcal{F}^{-1}[F(\mathbf{k})] = \int F(\mathbf{k}) e^{i2\pi \mathbf{k}\mathbf{r}}\ d\mathbf{k} = f(\mathbf{r}). \qquad (4)$$

Here $\mathbf{r}$ indicates real space 2D coordinates.

Without loss of generality, we can assume that the observed is a pure phase object with a weak phase and under weak phase assumption (WPA) object exit wave $\Psi(\mathbf{r})$ is described as

$$\Psi(\mathbf{r}) = exp(i\ \theta(\mathbf{r})) \cong 1 + i\ \theta(\mathbf{r}),\quad (\theta(\mathbf{r}) : \text{distribution of the projected phase of the object}). \qquad (5)$$

Since TEM images are formed through the double FT, the corresponding image intensity $I(\mathbf{r})$ and contrast $\Delta I(\mathbf{r})$ is given by the following formula.

$$I(r) = |\mathcal{F}^{-1}[\mathcal{F}[\Psi(r)] P(k)]|^2 = |\mathcal{F}^{-1}[\delta(k) P(k) + i\mathcal{F}[\theta(r)] P(k)]|^2$$
$$= |1 + i\theta(r) \otimes \mathcal{F}^{-1}[P(k)]|^2$$

$(\delta(k) P(k) = \delta(k) P(0) = \delta(k)$ assumed that reflects the optical axis positioning),

$$\cong 1 + 2R[i\theta(r) \otimes \mathcal{F}^{-1}[P(k)]], \quad (R[f(r)]\text{: real term extraction for } f(r))$$
$$= 1 - \sin\varphi(\theta(r) \otimes \mathcal{F}^{-1}[A(k)]) - ((1-\cos\varphi)/\pi x) \otimes \theta(r) \otimes \mathcal{F}^{-1}[A(k)] + O(\theta(r)^2). \tag{6}$$

$$\Delta I(r) = 1 - I(r) = -\sin\varphi(\theta(r) \otimes \mathcal{F}^{-1}[A(k)]) - ((1-\cos\varphi)/\pi x) \otimes \theta(r) \otimes \mathcal{F}^{-1}[A(k)]$$

(second order term in $I(r)$ ignored). (7)

In the above equation, $\otimes$ denotes the convolution of two functions and $|f(x)|$ the absolute value of $f(x)$. The first term *1* in Eq. 6 corresponds to the background due to the bright field derived from the un-scattered direct electrons, the second term to ZPC, the third term to HDPC, and the fourth term to the second order signal negligible under WPA. The symbol $(1/\pi x)\otimes$ appearing in the third term is implying the Hilbert transform in real space and the origin of the naming of HDPC. Note that only the HDPC term remains when the PP phase takes $\pi$, of which situation exactly represents the original HPP.

The results shown in Eqs. 6 and 7 looks troublesome from the practical viewpoint as two kinds of PC, ZPC and HDPC are mixed. However, this anomaly can be removed by two ways; one is to use an inverse filtering in spectral space [11] and the other is to use a two-experiment scheme using a pair of symmetric $\varphi$-HPPs (left and right) as claimed in this paper. The essential differences between the two will be detailed in Discussion. The right-sided $\varphi$-HPP is shown in Fig. 1c, where a half-planar $\varphi$-PP is placed in the right side contrarily to that shown in Fig. 1b. Due to this geometrical interchange, the exit wave modulator and the image contrast shown in Eqs. 2 and 7 are converted as

$$P(k) = 1/2\, A(k)[(1+\cos\varphi) - (1-\cos\varphi)\mathrm{sgn}(k_x) + i\sin\varphi(1+\mathrm{sgn}(k_x))], \tag{8}$$

$$\Delta I(r) = -\sin\varphi(\theta(r) \otimes \mathcal{F}^{-1}[A(k)]) + ((1-\cos\varphi)/\pi x) \otimes \theta(r) \otimes \mathcal{F}^{-1}[A(k)]. \tag{9}$$

From now on, in order to distinguish between the two modulator functions, we will refer to $P(k)$ in Eq.2 as $P_L(k)$ and $P(k)$ in Eq. 8 as $P_R(k)$.

The separation of the ZPC and HDC can be done by twin $\varphi$-HPPs, which combine paired symmetric images as obtained with a left-sided $\varphi$-HPP (refer Fig 1b and Eq. 7) and a right-sided $\varphi$-HPP (refer Fig 1c and Eq. 9).

Twin $\varphi$-HPPs sum contrast: $1 - 1/2\,(|\mathcal{F}^{-1}[\mathcal{F}[\Psi(r)] P_L(k)]|^2 + |\mathcal{F}^{-1}[\mathcal{F}[\Psi(r)] P_R(k)]|^2)$

$$= - sin\varphi( \theta (\boldsymbol{r}) \otimes \mathcal{F}^{-1}[A(\boldsymbol{k})]) \tag{10}$$

Twin $\varphi$-HPPs subtract contrast: $1/2( | \mathcal{F}^{-1}[ \mathcal{F}[\Psi(r)] P_L(\boldsymbol{k})] |^2 - | \mathcal{F}^{-1}[ \mathcal{F}[\Psi(r)] P_R(\boldsymbol{k})] |^2 )$

$$= ( (1- cos\varphi)/ \pi x ) \otimes \theta (\boldsymbol{r}) \otimes \mathcal{F}^{-1}[A(\boldsymbol{k})] \tag{11}$$

The HDPC shown in Eq. 11 can be converted into a ZPC by multiplying its spectral space counterpart ($-isgn(k_x)$) by a sign function ($isgn(k_x)$) and applying the relation ($sgn(k_x))^2 =1$. This is nothing other than the Hilbert transform in real space, and finally a highly sensitive ZPC image can be obtained by combining the two contrasts shown in Eqs. 10 and 11 as

Twin $\varphi$-HPPs unified contrast: Hilbert transformed twin $\varphi$-HPPs subtract contrast

$-$ twin $\varphi$-HPPs sum contrast

$$= (1- cos\varphi + sin\varphi)( \theta (\boldsymbol{r}) \otimes \mathcal{F}^{-1}[A(\boldsymbol{k})]) ,$$

$$= C_0(\varphi)( \theta (\boldsymbol{r}) \otimes \mathcal{F}^{-1}[A(\boldsymbol{k})] ) . \tag{12}$$

The index $C_0(\varphi)$ ($\equiv 1 - cos\varphi + sin\varphi$) in Eq. 12, which can be extended to CTFs in the next section, determines the sensitivity of the image contrast and takes following values at the characteristic PP phases,

$$C_0(\pi/2) = 2.00,\ C_0(11\pi/16) = 2.39,\ C_0(3\pi/4) = 2.41,\ C_0(7\pi/8) = 2.31 \text{ and } C_0(\pi) = 2.00. \tag{13}$$

Namely, the twin $\varphi$-HPPs unified contrast gives the strongest signal at $\varphi=3\pi/4$ with a sensitivity of 2.41, 20 % higher than 2, which is what can be obtained with conventional PP TEM with $\pi/2$-ZPP or $\pi$-HPP (refer $C_0(\pi)=2.00$) under the ideal experimental condition. Interestingly, $\pi/2$-HPP (refer $C_0(\pi/2) =2.00$) yield the same sensitivity of 2 as shown in Eq. 13. These findings unpredictable from our experience with conventional PPs completely rely on the novel PP scheme using twin $\varphi$-HPPs and an appropriate linear combination of the resultant image contrasts. We will delve into their background in Discussion. Here only mention that the $(1 - cos\varphi)$ term in $C_0(\varphi)$ ($\equiv 1 - cos\varphi + sin\varphi$) is rooted in HDPC and the $sin\varphi$ term in $C_0(\varphi)$ is in ZPC.

Fig. 1

*2.2 Optimum phase for $\varphi$-HPP under the condition of scattering electron dose loss*

Since those electrons transmitted through the object but scattered by PP become a uniform background with no contribution to the image formation, the relative dose of un-scattered electrons determines the transmission coefficient $T$ of PPs. The paper demonstrating high ZPC images of protein molecules [4a] was accompanied by a description of the transmission coefficient $T$ of $\pi/2$-PPs made of various materials at 100 kV and 300 kV acceleration voltages. According to the paper, for the amorphous carbon thin film, which is a typical material for PPs and assumed here, the coefficients are 0.62 for 100 kV, 0.72 for 300 kV and by interpolation 0.67 for 200 kV. Since these values are based on the number of the scattered electrons, they are considered to correspond to that of the intensity transmission. Therefore, it is reasonable to adopt $\sqrt{T}$ for the transmission coefficient of complex amplitude. In this case, $\sqrt{T}$ = 0,79 (100 kV), 0.82 (200 kV) and 0.85 (300 kV) for $\pi/2$-PPs. Using this amplitude transmission $\sqrt{T}$, the mathematical expression for the $\pi/2$-PP exit wave can be given by the following formula.

$\pi/2$-PP exit wave under the scattering electron dose loss:

$$\sqrt{T}\ exp(i\ \pi/2) = \sqrt{T}\ cos\pi/2 + i\ \sqrt{T}\ sin\pi/2. \qquad (14)$$

Associated with the above conversion from $exp(i\ \pi/2)$ to $\sqrt{T}\ exp(i\ \pi/2)$, the sensitivity index $C_0\ (\pi/2)$ must alter to $C_1(\pi/2)$ as

$$C_1(\pi/2) = 1 + \sqrt{T}\ sin\pi/2 - \sqrt{T}\ cos\pi/2. \qquad (15)$$

To extend Eq. 15 to cover an arbitrary phase $\varphi$, we can use the exponential dependence of the transmission on the PP thickness that is proportional to $\varphi$. By taking the $\pi/2$-PP coefficient as a standard, $C_1(\varphi)$ can be explicitly expressed with $\varphi$ as

$$C_1(\varphi) = 1 + T^{\varphi/\pi}\ sin\varphi - T^{\varphi/\pi}\ cos\varphi, \qquad (16)$$

where $T$= 0,62 (100 kV), 0.67(200 kV) and 0.72 (300 kV) for $\pi/2$-PPs of amorphous carbon. In Eqs. 15 and 16, the signal attenuation due to PP appears only in the second and third term.

Under the scattering electron dose loss condition, the sensitivity indices at three acceleration voltages of 100 kV, 200 kV and 300 kV become as follows.

For the case of 100 kV with $T = 0.62$ (for $\pi/2$ phase),

$$C_1(\pi/2) = 1.79,\ C_1(5\pi/8) = 1.97,\ C_1(11\pi/16) = 2.00,\ C_1(3\pi/4) = 1.99\ \text{and}\ C_1(\pi) = 1.62 \qquad (17)$$

For the case of 200 kV with $T = 0.67$ (for $\pi/2$ phase),

$$C_1(\pi/2) = 1.82,\ C_1(5\pi/8) = 2.02,\ C_1(11\pi/16) = 2.05_3,\ C_1(3\pi/4) = 2.04_6\ \text{and}\ C_1(\pi) = 1.67 \qquad (18)$$

For the case of 300 kV with $T = 0.72$ (for π/2 phase)

$$C_1(\pi/2) = 1.85, \ C_1(5\pi/8) = 2.06, \ C_1(11\pi/16) = 2.11, \ C_1(3\pi/4) = 2.10 \text{ and } C_1(\pi) = 1.72 \qquad (19)$$

It can be seen from above that the electron dose loss due to PP does not significantly change the optimum phase $\varphi = 3\pi/4$, which was obtained for $\varphi$-HPP under no electron dose loss condition.

*2.3 Optimum phase of φ-HPP under scattering and surface charge electron dose losses*

The surface charge of PPs is thought to be another source of the electron dose loss and often related to the detective quantum efficiency (DQE), which measures the statistical performance of radiation detectors originally defined as the quotient of the squared signal-to-noise ratio (SNR) at the output and input of the detector [18]. In general, it is defined in spectral space as a damping function. A recent paper [8] reported an extension of the DQE to cover a PP (VPP) made of amorphous carbon; for example, 0.94 at a wavenumber of 0.5 nm$^{-1}$ (corresponding to a resolution of 2 nm) and 0.4 for a wavenumber of 5 nm$^{-1}$ (corresponding to a resolution of 0.2 nm).

Let this DQE be expressed by a symbol, $D(k)$ ($k = |\mathbf{k}|$); that is, $D(0.5) = 0.94$ and $D(5) = 0.4$. Assuming that the two kinds of electron dose losses mentioned are multiplicatively superposed, the $\varphi$-PP exit wave can be expressed by the following formula.

$\varphi$-PP exit wave under the two kinds of electron dose losses: $\quad T^{\varphi/\pi} D(s) \exp(i\varphi)$. $\qquad (20)$

With above, the sensitivity index of Eq. 16 can be rewritten as

$$C_2(\varphi) = 1 + T^{\varphi/\pi} D(k) \sin\varphi - T^{\varphi/\pi} D(k) \cos\varphi. \qquad (21)$$

Then, the sensitivity indexes at two frequencies, $k = 0.5$ nm$^{-1}$ and 5 nm$^{-1}$, become as follows.

At $k = 0.5$ nm$^{-1}$ with $D(0.5) = 0.94$ for the case of 200 kV with $T = 0.67$ (for π/2 phase),

$$C_2(\pi/2) = 1.77, \ C_3(5\pi/8) = 1.96, \ C_2(11\pi/16) = 1.99, \ C_2(3\pi/4) = 1.98 \text{ and } C_2(\pi) = 1.63, \qquad (22)$$

At $k = 5$ nm$^{-1}$ with $D(5) = 0.4$ for the case of 200 kV with $T = 0.67$ (π/2 phase),

$$C_2(\pi/2) = 1.33, \ C_2(5\pi/8) = 1.41, \ C_2(11\pi/16) = 1.42_1, \ C_2(3\pi/4) = 1.41_9 \text{ and } C_3(\pi) = 1.27. \qquad (23)$$

Similarly, as $\varphi$-HPP, we can also define the sensitivity index for $\varphi$-ZPP as

$$C_{ZZ}(\varphi) = 2\, T^{\varphi/\pi} D(k) \sin\varphi. \qquad (24)$$

At $k$= 0.5 nm$^{-1}$ with $D(0.5) = 0.94$ for the case of 200 kV with $T = 0.67$ (for $\pi/2$ phase),

$C_{Z2}(\pi/2)$ =1.54, $C_{Z2}(5\pi/8)$ =1.35, $C_{Z2}(11\pi/16)$ =1.19, $C_{Z2}(3\pi/4)$ =0.98 and $C_{Z2}(\pi)$=0.　　　(25)

At $k$= 5 nm$^{-1}$ with $D(5) = 0.4$ for the case of 200 kV with $T = 0.67$ ($\pi/2$ phase),

$C_{Z2}(\pi/2)$=0.66, $C_{Z2}(5\pi/8)$ =0.58, $C_{Z2}(11\pi/16)$ =0.51, $C_{Z2}(3\pi/4)$ =0.42 and $C_{Z2}(\pi)$=0.　　　(26)

Since the surface charge electron dose loss is independent of the PP thickness, it is also independent of the PP phase $\varphi$. Therefore, the maximum sensitivity comes at $\varphi$ = 11π/16 as like for the case of scattering electron dose loss alone, but the amount itself is reduced due to the additional loss by the surface charge. Note that the sensitivity ratio between π/2-HPP and π/2-ZPP, especially in the high-frequency range at $k$ = 5 nm$^{-1}$ (corresponding to 0.2 nm resolution), is 1.33/0.66 = 2.02, a twofold improvement. The average sensitivity ratio in the low to high-frequency range (0.5 nm$^{-1}$ to 5 nm$^{-1}$) is 1.71 and the 71% improvement in sensitivity makes the 11π/16-HPP attractive.

**3. Contrast Transfer Theory to Evaluate the Sensitivity of Phase-contrast Methods**

Advantages and disadvantages of the two PPs, HPP and ZPP described in previous sections do not necessarily answer the crucial question about the PP method itself blitzed in Ref. 8 as the cause of performance degradation. We need to compare the sensitivity with the Scherzer defocus PC, which is free from PPs and proposed as a compromise of two aberrations, the defocus and spherical aberration, to maximize PC [6]. The contrast transfer theory originally developed in spectral space tells that the exit wave modulation has the form of $P(\boldsymbol{k})exp(-i\gamma(\boldsymbol{k}))$ instead of the simple form of $P(\boldsymbol{k})$ shown in Eq. 2. The phase disturbance function $exp(-i\gamma(\boldsymbol{k}))$, which represents the effect of lens aberrations represented by $\gamma(\boldsymbol{k})$, is most crucial to determine the image contrast. This is schematically shown in the following conversion of the image intensity shown in Eq. 6.

$$I(\boldsymbol{r}) = |\mathcal{F}^{-1}[\mathcal{F}[\Psi(\boldsymbol{r})]\,P(\boldsymbol{k})]|^2 \;\;\to\;\; I(\boldsymbol{r}) = |\mathcal{F}^{-1}[\mathcal{F}[\Psi(\boldsymbol{r})]\,P(\boldsymbol{k})\,exp(-i\gamma(\boldsymbol{k}))]|^2 \quad (27)$$

On the basis of above conversion and the associatively introduced new function of CTF, we need to investigate the advantages and disadvantages of the three PC methods (the definition of CTF is referred to Appendix). The theory of (phase) contrast transfer was developed to evaluate the influence of objective lens to imaging irrespective of other device factors such as TEM instabilities.

However, influence of TEM instabilities to contrast transfer as image damping was serious and they had been included to CTF as envelope functions [19]. To distinguish the extended CTF from the original one, we term it as the damped CTF as necessary.

Two enveloping functions are known. The envelope function $K_C(k)$ is characterized by the chromatic aberration and energy spread that reveals itself thorough device instabilities such as the instability of accelerating voltage and objective lens current (termed as chromatic aberration envelope function) [19]. The envelope function $K_S(k)$ is characterized by the opening angle of illumination aperture, which represents the partially coherence of the electron source (termed as source coherence envelope function) [19]. To make clear what is the origin of damping factors, the damped CTF can be explicit as 3-function product CTF (3FP-CTF) as $CTF(k) \cdot K_C(k) \cdot K_S(k)$, for example, as necessary.

Sensitivity comparison among different PC methods, the subject of this paper, was made using integrals of the image contrast in spectral space, specifically, using the integration from 0 to the cut-off frequency of the respective damped CTF. The cut-off frequency here is not that determined by the aperture but that at which the CTF first tends to zero, which is usually set to a lower one than the aperture cut-off frequency. We refer it as the CTF cut-off frequency. Next various functions referred so far are explicitly defined [20].

Lens aberration function: $\gamma(k) = 0.5\pi(2\Delta z \lambda k^2 + C_S \lambda^3 k^4)$ (28)

($C_S$: spherical aberration coefficient, $\lambda$: electron wavelength, $\Delta z$: amount of defocusing).

Here, every symbol takes a unit of length. There are two conventions about the sign of the defocus value $\Delta z$. Reimer defined positive $\Delta z$ to correspond the under-focus (object lens weakened) [21]. On the other hand, Nellist defined negative $\Delta z$ to correspond the under-focus [20]. We follow the Nellist convention according to the recent trend. In the under-focus condition ($\Delta z < 0$), the lens aberration function $\gamma(k)$ begins negative at lower frequencies, dominated by the first term in the right side of Eq. 28, and changes to positive at higher frequencies, where the second term dominated.

With the lens aberration function $\gamma(k)$, CTF for Scherzer defocus PC is given as [6b, 22]

CTF for Scherzer defocus PC: $2 \sin(-\gamma(k)) = -2 \sin \gamma(k)$. (29)

The negative sign in the above equation makes CTF positive at lower frequencies. Since $\sin \gamma (0) = 0$, this transfer function starts from zero with a severe suppression at lower frequencies. This means that the low-frequency components that determine the contrast of large structures are not transmitted to the image, resulting in an apparent low contrast. In Scherzer defocus, the balance between $C_S$ and $\Delta z$ is adjusted so that $2\sin \gamma (k)$ rises quickly at low frequencies and remains large at high frequencies. The corresponding defocus thus determined is called the Scherzer defocus and the corresponding resolution given as the zero-crossing point of the CTF called the Scherzer limit [6a] or here called the CTF resolution, which is defined for example by $0.625 (C_s \lambda)^{1/4}$ [6b]. This makes the Scherzer defocus PC popular in various fields relying on TEM, particularly in material sciences. Note that Eq. 29 represents a general formula not restricted to the case of the Scherzer defocus but termed here as Scherzer defocus PC to characterize the high sensitivity feature of such a defocus setting.

Th CTF for $\varphi$-ZPP PC can be obtained by changing $-\gamma (k)$ in Eq. 29 to $-\gamma (k) + \varphi$ and adding the electron dose loss effect as shown in Eq. 20.

CTF for $\varphi$-ZPP PC:

$$2T^{\varphi/\pi}D(k)\sin(-\gamma (k) + \varphi) = 2T^{\varphi/\pi}D(k)(\sin\varphi\cos \gamma (k) - \cos\varphi\sin \gamma (k)) \quad (30)$$

For $\varphi = \pi/2$, Eq. 30 reduces to $2T^{\varphi/\pi}D(k) \cos \gamma (k)$ and low-frequency components are well transferred, although an intensity reduction due to the PP electron dose loss advents as the factor $T^{\varphi/\pi}D(k)$. The low-frequency components can add a high contrast appearance to the TEM image but as will be shown later, the integral intensity of the PC seen over the whole frequency range is not necessarily high. In the same way as for the Scherzer defocus method, $\Delta z$ can be adjusted so that $\cos \gamma (k)$ remains large up to higher frequencies [7b].

Th CTF for twin $\varphi$-HPPs unified PC is given below (refer appendix for the derivation).

CTF for twin $\varphi$-HPPs unified PC:

$$\sqrt{2} (\cos(-\gamma (k) - \pi/4) + T^{\varphi/\pi}D(k)\sin(-\gamma (k) + \varphi - \pi/4))$$

$$= (1 + T^{\varphi/\pi}D(k) \sin\varphi - T^{\varphi/\pi}D(k) \cos\varphi) \cos \gamma (k)$$

$$- ( 1 + T^{\varphi/\pi}D(\mathbf{k}) \sin\varphi + T^{\varphi/\pi}D(\mathbf{k}) \cos\varphi ) \sin\gamma(\mathbf{k}). \quad (31)$$

From the comparison between Eqs. 21 and 31, we find that Eq. 31 (CTF for twin $\varphi$-HPPs unified PC) converges to Eq. 21 ($C_2(\varphi)$ sensitivity index) when $\gamma(\mathbf{k})=0$. This is rather obvious as no aberration assumed to derive Eq. 21 means $\gamma(\mathbf{k})=0$. From other angles, CTF for twin $\varphi$-HPPs unified PC can be seen as a splitting of $C_2(\varphi)$ sensitivity index to two terms respectively with the modulation of $\cos\gamma(\mathbf{k})$ or $\sin\gamma(\mathbf{k})$. This symbolizes a natural extension from sensitivity index to CTF.

If we set $\pi/2$ to $\varphi$, Eq. 31 takes a simple form of $( 1 + T^{\varphi/\pi}D(\mathbf{k}))(\cos\gamma(\mathbf{k}) - \sin\gamma(\mathbf{k}))$. The addition of the two CTFs, $\cos\gamma(\mathbf{k})$ corresponding to the $\pi/2$-ZPP CTF and $-\sin\gamma(\mathbf{k})$ to the Scherzer defocus CTF, makes the twin $\varphi$-HPPs unified CTF unique. The unique form is to be examined from the viewpoint of COBS in Discussion.

Next, we will define envelope functions [19].

Chromatic aberration envelope function: $K_C(\mathbf{k}) = exp[-(\pi\lambda H/(4\sqrt{\ln 2}))^2 \mathbf{k}^4]$ (32)

Above, $H$ is a parameter with a length unit (termed as reduced chromatic aberration) dependent on chromatic aberration $C_c$, energy spread $\Delta E$ that originates from the lens current instability and the accelerating voltage instability, and the accelerating voltage $E$. ln2 is the natural logarithm of 2, which is about 0.693.

Source coherence envelope function: $K_S(\mathbf{k}) = exp[-(\pi^2\alpha^2(C_S^2 \mathbf{k}^2 - \Delta z)^2/\ln 2)\mathbf{k}^2]$ (33)

Above, $\alpha$ is the opening angle of illumination aperture in radians. The others have already been mentioned.

Next, DQE is defined according to the report [8].

Detection quantum efficiency (DQE): $D(k) = 1 - 0.12s$ (34)

The above functional form is a linear approximation of the DQE for the case of the thinnest PP (3 nm thick) (for example, $D(5) = 0.4$ (60% loss)).

## 4. Visual Representation of CTFs and Sensitivity-Evaluation Integrals

CTF is usually illustrated as a 1D function $CTF(k)$ with radial coordinates, on the ground that the functions involved preserve a rotational symmetry. Specific examples are shown in Fig.2a, where the three kinds of CTFs corresponding to the case, in which a combination of $\varphi$ and $\Delta z$ gives maximum sensitivity for the twin $\varphi$-HPPs unified PC and $\Delta z$ gives maximum sensitivity for the Scherzer focus and $\pi/2$-ZPP PC, are displayed assuming other experimental parameters fixed as given below in Eq. 35. Details of the maximum sensitivity search are to be explained later.

$\lambda$ = 2.5x10-3 nm (200 kV acceleration voltage), $C_S$ = 0.4 mm, $\alpha$ = 1 mm rad, $H$ = 5 nm.    (35)

Twin $\varphi$-HPPs unified PC CTF: $\varphi$ = 1.29 rad, $\Delta z$ = -31 nm, $k_C$ = 5.2 nm$^{-1}$ in the ($\varphi$, $\Delta z$) search.  (36)

Scherzer defocus PC CTF: $\varphi$-independent, $\Delta z$ = -38 nm, $k_C$ = 5.5 nm$^{-1}$ in the $\Delta z$ search.    (37)

$\pi/2$-ZPP PC CTF: $\varphi$ =1.57 ($\pi/2$) rad, $\Delta z$ = -21 nm, $k_C$ = 4.8 nm$^{-1}$ in the $\Delta z$ search.    (38)

Here $k_C$ is the CTF cut-off frequency that corresponds to the Scherzer limit or CTF resolution as a guideline for the spatial resolution; for the cases of Eqs. 36 to 38, 0.18 nm for Scherzer defocus PC, 0.19 nm for twin $\varphi$-HPPs unified PC and 2.1 nm for $\pi/2$-ZPP PC. Incidentally, the CTF resolution obtained from the Scherzer limit formula described earlier was 0.176 nm, which is naturally the same as 0.18 nm for Scherzer defocus PC.

In Fig. 2a, to emphasize the significance of the CTF cut-off frequency that was used in the sensitivity estimation as the integral limit, the frequency range to cover the information limit is cut. But just to avoid misunderstanding, we would like to add that the spatial resolution defined by the information limit that is unshown in Fig. 2 is much higher under the experimental parameters given in Eq. 35, for example, 0.12 nm for Scherzer defocus PC.

In Fig. 2a, we can recognize the other features characteristic to the two CTFs, the Scherzer focus PC (red line) and $\pi/2$-ZPP PC (blue line) CTF, the former as a high pass filter and the latter as a low pass filter. On the other hand, the twin $\varphi$-HPPs PC CTF (magenta line) looks as an intermediate of the two, namely covering the range from 0 to 5 nm$^{-1}$ almost in a flat feature. Moreover, its height exceeds both conventional CTFs also in the range, inferring its high sensitivity.

Fig. 2

To assess sensitivity particularly with use of the CTF-based sensitivity, the damped CTF or 3FP-CTF (termed as $C_3(\varphi;\boldsymbol{k})$) can be used. With this new sensitivity index, the CTF-based sensitivity can be defined as an integral in 2D spectral space as Sensitivity integral: $\bar{\bar{C}}_3(\varphi) = \iint_0^{k_C} C_3(\varphi; \boldsymbol{k}) d\boldsymbol{k} = 2\pi \int_0^{k_C} C_3(\varphi; k) k\, dk$ , (39) where, $k_C$ is the CTF cut-off frequency. Using the rotational symmetry of $C_3(\varphi;k)$, the 2D integration is reduced to that of 1D one in Eq. 39. The 4-function product $C_3(\varphi; k)k$ or $kC_3(\varphi; k)$, instead of the 3-function product $C_3(\varphi; k)$, is one of core messages in this study.

Three damped 3FP-CTFs for twin $\varphi$-HPPs unified PC, the Scherzer defocus PC and the $\varphi$-ZPP PC are given as

$C_{H3}(\varphi; k) = \sqrt{2}\,(cos(\gamma(k) - \pi/4) + T^{\varphi/\pi} D(s) sin(\gamma(k) + \varphi - \pi/4))\, K_C(k) K_S(k)$ , (40)

$C_{S3}(k) = -2 sin\,\gamma(k) K_C(k) K_S(k)$, (41)

$C_{Z3}(\varphi;k) = 2 T^{\varphi/\pi} D(s)\, sin(-\gamma(k) + \varphi)\, K_C(k) K_S(k)$ . (42)

Note that the $C_{S4}(k)$ is irrelevant to $\varphi$, as the Scherzer defocus is free from PP. The CTFs shown in Fig. 2 are those corresponding to 3FP-CTFs $C_3(\varphi; k)$ (a) and 4FP-CTFs $kC_3(\varphi; k)$ (b). Let us explain the core assertion next,

The sensitivity maximum was searched using the sensitivity integral $\bar{\bar{C}}_3(\varphi)$ given in Eq. 39 for each of three PC CTFs given in Eq. 40 to Eq. 42. The search was carried out with the Δz scanning for each of φ for the twin φ-HPPs and φ-ZPP integration, and for Scherzer defocus integration with use of the same parameters shown in Eq. 35. The particular 3FP-CTFs and 4FP-CTFs corresponding to the sensitivity maximum thus found are shown in Figs. 2a and 2b respectively for three PC methods. It needn't be said but the sensitivity integral of Eq. 39 was carried out for the 4FP-CTFs shown in Fig. 2b.

A comparison of 3FP-CTFs (Fig. 2a) and 4FP-CTFs (Fig. 2b) reveals a feature of the Scherzer defocus method that has not attracted much attention so far. That is, in 4FP-CTFs, the additional product by *k* mitigates the relative disadvantage in the low-frequency range of the Scherzer defocus PC. Conversely, it emphases the advantage in the high-frequency range. The low-frequency advantage of the π/2-ZPP PC is cancelled out and rather transforms into disadvantage. On the other hand, twin φ-HPPs unified PC, due to the feature of intermediate between Scherzer defocus and π/2-ZPP, appears to be superior to Scherzer defocus PC both in 3FP-CTF and in 4FP-CTF and retain the high sensitivity. With use of a measure of relative quantities, say sensitivity ratio, this will be illustrated next.

The four charts in Fig. 3 with varying *H* from 5 nm to 100 nm illustrate sensitivity ratios between twin φ-HPPs unified PC vs Scherzer defocus PC (refer magenta line in Fig. 3) and π/2-ZPP PC vs

Scherzer defocus PC (refer blue line in Fig. 3). Comparative overview of the four charts immediately reveals three features; first the sensitivity ratio for HPP vs Scherzer is always far better than that for ZPP vs Scherzer, second the larger $H$ becomes, bigger the two sensitivity ratios are and third as $H$ increases, the discrepancy between theory and simulation increases. The first and the second seem to visually reproduce what has already been pointed out in the previous section as the $\varphi$-dependence. The third is to be discussed in the section followed. Next, the characteristics of the three PCs in terms of specific values of sensitivity ratios are compared.

Two kinds of sensitivity ratios between twin $\varphi$-HPPs vs Scherzer and $\varphi$-ZPP vs Scherzer but twin $\varphi$-HPPs vs $\varphi$-ZPP are shown in Fig. 3. This is because the values for twin $\varphi$-HPPs/$\varphi$-ZPP sensitivity ratios do not fit into Fig. 3, as they become too large towards $\varphi = \pi$. Nevertheless, the sensitivity ratios of the three kinds, including the twin $\varphi$-HPPs/$\varphi$-ZPP sensitivity ratio, for the case of $\varphi = \pi/2$, which is of particular interest in comparison with PC methods proposed and used in the past, are given below.

$H=5$;
twin $\pi/2$-HPPs/Scherzer: 1.02, twin $\pi/2$-HPPs/$\pi/2$-ZPP: 2.39, $\pi/2$-ZPP/Scherzer: 0.43,  (43)

$H=20$;
twin $\pi/2$-HPPs/Scherzer: 1.16, twin $\pi/2$-HPPs/$\pi/2$-ZPP: 2.15, $\pi/2$-ZPP/Scherzer: 0.54,  (44)

$H=40$;
twin $\pi/2$-HPPs/Scherzer: 1.42, twin $\pi/2$-HPPs/$\pi/2$-ZPP: 1.85, $\pi/2$-ZPP/Scherzer: 0.77,  (45)

$H=100$;
twin $\pi/2$-HPPs/Scherzer: 1.72, twin $\pi/2$-HPPs/$\pi/2$-ZPP: 1.58, $\pi/2$-ZPP/Scherzer: 1.09,  (46)

By presenting more specific values as shown above, new findings and reaffirmation of old findings can be seen more clearly: the superiority of twin $\pi/2$-HPPs over Scherzer is higher on the low-resolution side (larger H) than on the high-resolution side (smaller H). Rather, the comparable sensitivity of Scherzer to twin $\pi/2$-HPPs at high-resolution H=5 is worth mentioning. On the other hand, the superiority of twin $\pi/2$-HPPs over $\pi/2$-ZPP is higher on the high-resolution side (smaller H) than on the low-resolution side (larger H). Moreover, the difference is remarkable under all conditions and the ratio of 2.39 at high-resolution H=5 is particularly noteworthy. Finally, the sensitivity of $\pi/2$-ZPP is far less than that of Scherzer on the high-resolution side (smaller H) and $\pi/2$-ZPP is only slightly superior to Scherzer on the low-resolution end H=100. In a sense, this reproduces an old finding, namely the expectation of a contrast-enhancing effect for the $\pi/2$-ZPP method. The implications of these results will be discussed in more detail in the Discussion.

Through the φ-dependence shown in Fig. 3, we can recognize the other characteristics proper to the PP methods. For both PP methods, the PP phase that gives maximum value of the sensitivity ratio increases and moves towards the large phase side while $H$ increases, namely when lowering the spatial resolution of images. For twin φ-HPPs unified PC, the range of movement is between 6π/16 and 9π/16 centered on π/2 and, for φ-ZPP PC, between 0 and 6π/16 centered on 3π/16. While the result for twin φ-HPPs unified PC reaffirms the assertions made in the previous section based on Eq. 31, the result for the φ-ZPP PC is an unexpected surprise, because it means that π/2 ZPP is not the best among φ-ZPPs but a φ-ZPP with smaller phase φ is better. The meaning of these results is also to be discussed in Discussion.

Fig. 3

## 5. A Verification of the High Sensitivity of Twin φ-HPPs PC with a TEM Simulator

Experimental verification is needed to confirm the theoretical results described so far. Achieving this with real TEM equipment is, however, rather difficult because of the enormous cost and time required. Fortunately, this difficulty can be overcome in TEM by employing a TEM simulator. TEM experiments can be mimicked with computational means as electron optical processes behind a TEM experiment can be exactly reproduced with use of firmly established physical principles of electron scattering phenomena, as long as the atomic structure of the object under investigation is known. This is the reason why numerous references were sought from existing textbooks and handbooks and why a TEM simulator or virtual TEM, which could simulate the TEM imaging process in a high speed and with an accuracy at will, was used in this study.

The TEM simulator used was Elbis™ from FH Electron Optics in Japan [https://www.fhelectronopt.com/]. Elbis™ is a comprehensive simulator with a wide range of applicability comprising of numerical calculations of multilayer electron scattering and multiple integrations of TCC (Transmission Cross Coefficient), the main feature of which is that the numerical calculations are processed in parallel using a GPU (Graphics Processing Unit), thus achieving high speed. Currently, about 50 sets are in use in Japan. Details are given in the paper of benchmark tests [23].

CTF theory concerns the instrument performance, meaning no concern to the characteristics of the object under investigation. However, real experiments and simulations as well require specific objects for observation. In this paper, a single gold atom was employed as the object simulated. This is because the atomic scattering factor of heavy atoms seems to be flat up to around 5 nm$^{-1}$ [24] and the intensity and spread of gold image is large enough to facilitate visualization. In Fig. 3, results for HPP vs Scherzer sensitivity ratio of the simulated gold atom images are also shown with red circles in addition to those evaluated through the theory. The simulation details are illustrated in Sect. 6 together with the simulated images.

As mentioned before, as the reduced chromatic aberration $H$ increases, the discrepancy between theory (magenta diamonds) and simulation (red circles) increases. Currently the reason for the trend is not clear but it is not unfair to say that the superiority of twin $\varphi$-HPPs PC has been demonstrated both theoretically and by simulation. Specifically, the following was observed. For $H$= 5 nm, the maximum sensitivity ratio is 1.03 (for theory at $\varphi=3\pi/8$) and 1.10 (for simulation at $\varphi=3\pi/8$), for $H$= 20 nm, the maximum sensitivity ratio is 1.16 (for theory at $\varphi=7\pi/16$) and 1.16 (for simulation at $\varphi=3\pi/8$), for $H$= 40 nm, the maximum sensitivity ratio is 1.42 (for theory at $\varphi=\pi/2$) and 1.28 (for simulation at $\varphi=\pi/2$) and for $H$= 100 nm, the maximum sensitivity ratio is 1.73 (for theory at $\varphi=9\pi/16$) and 1.38 (for simulation at $\varphi=\pi/2$). Thus, even though there are some differences in the results between theory and simulation, the superiority of the twin $\varphi$-HPPs method can be claimed irrespective of the variation of reduced chromatic aberration $H$. On the other hand, the ZPP method could not overcome negative aspects of the PP, i.e. electron dose losses, even though the PP phase was varied. In relation to electron dose losses, an asymmetry between theory and simulation exists, which is to be referred next.

In the upgraded Elbis™ that can facilitate the PP function, among two features of the electron dose loss, only the scattering loss was included but not the DQE-related charging effect. This is because there was no established theory on the relationship between the charging effect and DQE. Therefore, we included the DQE effect into the CTF formulation as a phenomenological equation (refer Eq. 34) but dared to avoid the inclusion in the simulator. The gap between the theory and simulation was filled in the simulation side by substituting the damping effect of DQE by that of chromatic aberration envelope function $K_C(k)$. The substitution was simple: just adjust the reduced chromatic aberration $H$ of $K_C(k)$ to a larger value, if DQE= 1.

To estimate the enlarged $H$ adequate for the simulation to have the substituted damping effect, the integral equivalence between the two 4FP-CTFs, one used in the theory including DQE and the other to appear in the simulation without DQE, was used for the $H$ adjustment. Examples of adjusted $H$s are shown in Fig. 4 together with the original $H$s. The DQE has a greater attenuation effect on the high-frequency side, so the difference between the two CTFs stands out for the case of smaller $H$ corresponding to the high-frequency coverage. Taking this into account and comparing Figs. 3 and 4, we find that the correspondence between the theoretical (diamonds in Fig. 3) and the simulated (circles in Fig. 3) is more similar for smaller $H$ (5 and 20 nm) but divergent for larger $H$ (40 and 100 nm). The former implies the adequacy of the integral equivalence for the substitution criterion of $H$ adjustment, while the latter is considered to mean that the deviations do not necessarily originate from the substitution method itself. The reasons for the divergence itself will be reconsidered in Discussion.

Fig. 4

## 6. Images Simulated for a Single Atom of Gold and Experimental Flow of the Twin $\varphi$-HPPs Method

The physical basis of the disparity seen in the comparison between CTF theory and simulation, the former disregarding the object but the latter including it, must be clarified. While images of a single gold atom are simulated in the present study, the exact solution of the image contrast of a single atom based on electron-atom collision theory is already known as [25],

$$\Delta I(r; \varphi) = 4\pi \int_0^{k_o} f(k) C_4(\varphi; k) J_0(2\pi kr) k \, dk, \tag{47}$$

where $k_0$ is the aperture cut-off frequency, $f(k)$ the electron scattering amplitude of an atom, $C_4(\varphi; k)$ the damped CTF including the envelope function (original expression of the CTF [25] was $\sin\gamma(\mathbf{k})$ but here extended to cover those PP systems show in Eqs. 40 and 42), $J_0(2\pi kr)$ is the zero-order Bessel function and $J_0(0) = 1$. Equation 47 was given under WPA, and heavy metals were considered out of scope. However, since the phase shift induced by a gold foil of one atom thickness at 200 kV is less than 1 mm rad, one single gold atom is also considered belonging a weak phase object, and the application of the exact solution Eq. 47 to a gold atom can be allowed.

Considering $J_0(0) = 1$ at the center of the gold atom ($r=0$) and $f(k) = f(0)$ that reflects the fact that electron scattering amplitude $f(k)$ of heavy atoms such as gold atom is fairly flat up to a frequency of 5 nm$^{-1}$ [24], Eq. 47 is simplified at $r=0$ as,

$$\Delta I(0; \varphi) = 4\pi f(0) \int_0^{k_o} C_4(\varphi; k) k \, dk. \tag{48}$$

Apart from the constant coefficient, above Eq. 48 can be attributed to the sensitivity integral Eq. 39. This is a justification for the contrast comparison between the CTF theory, which needs not to specify the object, and the simulation, which needs to specify the object.

By the way, the cut-offs of the integration edge in the two equations (Eq. 39 and Eq. 48) have different physical contents, but we have made them consistent by aligning the aperture cut-off in Eq. 48 with that in Eq. 39. In other words, the CTF cut-off was applied to the aperture in the simulation.

With above considerations keeping in mind, let us see the simulation results for a single gold atom. Figure 5 shows simulations corresponding to the high-resolution experimental setting, namely $H=5$ and Fig. 6 shows simulations corresponding to the low-resolution experimental setting, namely $H=100$. Both figures illustrate the data flow leading to the final image (twin $\varphi$-HPPs unified image). As high-resolution images have sharper features that are easy to follow, we will focus on Fig. 5 that illustrates more details, for example, the mixed feature of the normal and differential PC in the most upper row.

The start of the data flow is the most upper row in Fig. 5, where the twin images of a gold atom for the 1.29-HPP are shown. A pair of two images were yielded with successive two simulations only displacing the optical axis from left to right or vice versa for the 1.29-HPP, of which setup is shown in Fig. 1d. For the optical axis displacement, experimentally, both of the mechanical and beam deflection method are possible, the latter being preferred in terms of high speed [26]. Note that both images are a mixture of the normal-type and differential-type image, namely skewed images, where the direction of the skewness is reversed each other.

The second row illustrates sum and subtract of the twin images, where the sum corresponds to a normal-type (or ZPC type) and the subtract to a differential-type (or HDPC type). The difference of the two images is not only clear from their different appearance but also the difference in the brightness of their background. The background of each of original twin images is *1* as described so far and that of the sum is also *1* after dose normalization (namely divided by 2). On the other hand, the background of the subtract vanishes and the differential image can take positive and negative values along the differential direction. Usually, images are displayed with a gray tone taking values smaller than the background *1* but larger than 0. To compromise the negative value problem in the differential image, a small positive constant is added to the differential image to make the whole positive. This is the image shown in the right side of the second row.

Third step is the Hilbert transform (HT) of the subtract image to recover the normal-type PC from the differential-type one as shown in the right side of the third row. Here, the background problem mentioned disappears.

The unification of the two images, the sum and Hilbert transformed subtract image, illustrates a highly sensitive image as shown in the center of the bottom row. Note that dose normalization is not required in the unification as the two images are independent. In the bottom row, for the comparison, two PC images taken with the Scherzer defocus (left side in the bottom of Fig. 5) and the $\pi/2$-ZPP (right side in the bottom of Fig. 5) are shown side by side. It is clear that the twin 1.29-HPPs unified image is better in contrast against the other two images, particularly better by about 2 against the $\pi/2$-ZPP image. The twofold increase in sensitivity is very impressive and shows the future potential of the new HPP method.

Fig. 5

The data flow in Fig. 6, which shows the simulation results with the 1.80-HPP for $H$=100, is same as for the case of Fig. 5 and the details are omitted. But four features specific to this case can be pointed out; (1) broadening, (2) intensity reduction, (3) the shift of the sensitivity maximum phase from $3\pi/8$ ($H$ = 5) to $\pi/2$ ($H$ = 100) and (4) the sensitivity of the twin $\varphi$-HPPs unified PC relative to the Scherzer defocus PC being larger for $H$ = 5 than for $H$ = 100. Features (1) and (2) are natural because the signal attenuation effect due to the envelope function is larger for $H$ = 100 than for $H$ = 5. For the feature (3), the considerations given in Section 5 may apply. For the feature (4), exactly an opposite reason given for (3) looks happening. In other words, at low-resolutions where low frequencies are emphasized, the Scherzer defocus PC method, which is disadvantageous for the low-frequency gain, becomes relatively less sensitive, and as a result the sensitivity of the twin $\varphi$-HPPs unified PC relative to the Scherzer defocus PC improves in the case of higher $H$. In comparison, as the $\pi/2$-ZPP PC has an advantage in the low-frequency gain, there is no significant difference in the sensitivity of the twin $\varphi$-HPPs unified PC relative to the $\pi/2$-ZPP PC between for $H = 5$ and for $H=100$.

In any case, the results presented in Figs. 5 and 6 demonstrate that the twin $\varphi$-HPPs unified PC has the greatest sensitivity among the three PCs investigated.

Fig. 6

## 7. Discussion

Since the image contrast severely depends on the setup of physical parameters and their values used in the specific investigations, types of physical parameters have to be the first thing to be clarified in the search of maximum contrast sensitivity. Following physical parameters in TEM are central; $\lambda$: wave length determined by the acceleration voltage $E$, $\Delta E$: energy spread of the acceleration voltage, $C_s$: spherical aberration, $C_c$: chromatic aberration, $H$: reduced chromatic aberration determined by $C_c$, $\Delta E$ and $E$, $\varphi$: PP phase, $T$: PP transmission coefficient, $\Delta z$: defocus aberration, $k_o$: aperture cut-off frequency, $k_C$: CTF cut-off frequency, In this paper, parameter values for $\lambda(E)$, $\Delta E$ and $C_s$ are fixed to those as normally used in TEM experiments, while parameter values for $\varphi$, $T$, $\Delta z$, $C_c$ (through $H$) $k_o$, and $k_C$ are varied according to the concern of the investigation, particularly concerning to the contrast sensitivity of PC methods. This allows the sensitivity comparisons among different PC methods within the parameter values specified. Nevertheless, we believe that the sensitivity comparisons developed here can be extended to cover a wider range of parameter values, if they do not significantly deviate from those specified here.

In TEM, the real space image is always real, but the corresponding spectral image can be complex: the mixed image seen in $\varphi$-HPP PCs ($\varphi < \pi$) corresponds to this case, and in the case of $\varphi = \pi/2$, it takes following complex-number contrast in spectral space as shown by Koeck [11] (note : $\pi/2$-HPP changed from the right sided to the left-sided).

$$\mathcal{F}[\Delta I(r)] = (1 - i\, sgn(k_x))(\cos\gamma(\mathbf{k}) - \sin\gamma(\mathbf{k}))\mathcal{F}[\Psi(r)] \qquad (49)$$

$$= CTF(\mathbf{k})\,\mathcal{F}[\Psi(r)]$$

The first term *1* in （*1 - i sgn(k$_x$)*) corresponds to the normal-type image and the second one to the differential-type image, and $\Delta I(r)$ gives a mixed form of the normal and differential real image. The complex-number $CTF(\mathbf{k})$ $(= (1 - i\, sgn(k_x)) (\cos\gamma(\mathbf{k}) - \sin\gamma(\mathbf{k})))$ can be converted to a real-number one through an inverse filter of $(1 + i\, sgn(k_x))$ as

$$\text{Realized } CTF(\mathbf{k}) = \sqrt{2}\,(\cos\gamma(\mathbf{k}) - \sin\gamma(\mathbf{k})), \tag{50}$$

where $\sqrt{2}$ arises from weighting for the inverse filter. The lens aberration part in Eq. 50 is isomorphic to the CTF shown in Eq. 31 for twin $\varphi$-HPPs unified PC when $\varphi = \pi/2$. Under the same condition of no electron dose loss as taken by Koeck, the corresponding $CTF(\mathbf{k})$ for the two-experiment scheme becomes

$$C_{H3}(\pi/2;\mathbf{k}) = 2\,(\cos\gamma(\mathbf{k}) - \sin\gamma(\mathbf{k})). \tag{51}$$

Comparing the coefficients of Eqs. 50 and 51, the twin $\pi/2$-HPPs unified PC is clearly higher in sensitivity by $\sqrt{2}$ than the Koeck right-sided $\pi/2$-HPP PC.

The Koeck approach seems to improve the spatial resolution but not necessarily the sensitivity. So next we must ask why such a discrepancy in sensitivity between the two approaches has occurred? Of course, it stems from the approach difference. Namely, Koeck approach is based on one-experiment scheme directly to recover the normal image using left-sided or right-sided HPP and an inverse filter. On the other hand, this report approach relies on two-experiment scheme using a pair of symmetric HPPs and a linear combination of resultant PCs. Actually, the inverse filter employed by Koeck is mathematically equivalent to the linear combination this report employed, if the Koeck's one experiment is virtually divided to two experiments and they are linearly combined. The signal part of the virtual two-experiment scheme must give the same result as that of the real two-experiment scheme but the noise part of the virtual two-experiment scheme differs to that of the real two-experiment scheme as noise becomes coherent in the former case and incoherent in the latter case. It is easy to understand the reason of improved signal-to-noise ratio associated with statistical averaging, which is only allowable to apply to the real two-experiment scheme, namely two experiments are statistically averaged and gain $\sqrt{2}$ improvement of the signal-to-noise ratio or $\sqrt{2}$ sensitivity enhancement.

Incidentally, two-time experiments are the essence of COBS [13, 14], and its correspondence to the twin π/2-HPPs method, which definitely needs two-time experiments, is to be briefly reviewed next together with what COBS implies.

COBS is a method to remove the effect of lens aberrations or CTF modulations from microscopic images or PCs through an inverse operation (demodulation) of the modulated exit wave [13]. The method is based on two successive TEM observations to recover the exit wave as its original form of a complex amplitude. Its distinctive feature is that the zero division inevitably manifested in the demodulation for the real intensity image as like $(sin\ \gamma\ (\mathbf{k}))^{-1}$ or $(cos\ \gamma\ (\mathbf{k}))^{-1}$. COBS can avoid the difficulty with a complex demodulator $exp(i\ \gamma\ (\mathbf{k})$ for the modulated form shown in Eq. 27, which is schematically shown below.

Demodulated $I\ (\mathbf{r}) = |\ \mathcal{F}^{-1}[\ \mathcal{F}[\Psi(\mathbf{r})]P(\mathbf{k})\ exp(-i\ \gamma\ (\mathbf{k}))\ exp(i\ \gamma\ (\mathbf{k}))\ ]\ |^{\ 2} = |\ \mathcal{F}^{-1}[\ \mathcal{F}[\Psi(\mathbf{r})]P(\mathbf{k})]\ |^{\ 2}$ (52)

The inverse operation to remove $exp(-i\ \gamma\ (\mathbf{k}))$ shown above is mathematically straightforward as

$exp(-i\ \gamma\ (\mathbf{k}))\ exp(i\ \gamma\ (\mathbf{k})) = (cos\text{-}\ \gamma\ (\mathbf{k}))^{2} + (sin\text{-}\ \gamma\ (\mathbf{k}))^{2} = = (cos\ \gamma\ (\mathbf{k}))^{2} + (sin\ \gamma\ (\mathbf{k}))^{2} = 1.$ (53)

In the real space, however, this cannot be performed, since images are obtained as real-valued entities in the detection plane. It can only be performed in the computer space for complex-number images reconstituted through the complex combination of the two real images separately obtained with defocusing and ZPP [13].

In connection with Eq. 31, it can be argued that the CTF takes the following form when $\varphi = \pi/2$,

$L(\mathbf{k})\ (cos\ \gamma\ (\mathbf{k}) - sin\ \gamma\ (\mathbf{k})) = L(\mathbf{k})\ (cos(-\ \gamma\ (\mathbf{k})\ ) + sin(-\ \gamma\ (\mathbf{k})),$

($L(\mathbf{k})$: electron dose loss coefficient).    (54)

Comparing Eqs.53 and Eq.54, it can be seen that the summation of the sine and cosine term is performed in the square form in Eq. 53 but done in a linear form in Eq. 54. To differentiate the two, therefore, the former is to be called second-order COBS and the latter first-order COBS. Since the first-order COBS $(cos(-\ \gamma\ (\mathbf{k})\ ) + sin(-\ \gamma\ (\mathbf{k})\ ))$ in Eq. 54 is almost constant in the range $0 < \gamma\ (\mathbf{k}) < \pi/2$, similarly to $(cos(-\ \gamma\ (\mathbf{k}))\ )^{2} + (sin(-\ \gamma\ (\mathbf{k}))\ )^{2} = 1$ in the second-order COBS, it could exclude both reduction of resolution and sensitivity due to lens aberrations. However, because of the incidental condition of $0 < \gamma(k) < \pi/2$ required for the first-order COBS, such a higher-resolution claimed by the second-order COBS cannot be attained. In any case, the most important feature of COBS is to eliminate the effects of aberrations entirely, whether or not it is first- or second order. In other words, as long as the aberration function $\gamma(\mathbf{k})$ is an even function, it makes the existence of aberrations silent, regardless of its details going through lower to higher order aberrations.

The original second-order COBS requires an asymmetric pair of PC methods of defocusing and ZPP. In contrast, the twin $\varphi$-HPPs unified PC requires a symmetric pair of HPPs. As a consequence of this difference, for the first-order COBS, it is not infeasible to simplify the two-experiment scheme to a one-experiment one, while it is a pseudo-kind. An example of such a pseudo one-experiment scheme is given in Ref. 26.

The difference between apparent contrast and sensitivity contrast defined by the sensitivity integral, already foretold in Note 1, is to be discussed from the perspective of imaging methods that aim to faithfully reproduce the structure of objects.

Danev (the developer of VPP) and his team performed single-particle analysis using small membrane proteins (G protein-coupled receptors: GPCRs) and discussed the difference in resolution achieved with and without VPP [27], as a sequel to the DQE paper [8] that warned against the PP method. In support of the DQE paper, the results with VPP failed to clear the 0.3 nm resolution, while the results without VPP combined with deep defocusing and CFT correction came close to 0.25 nm resolution. At first glance, these results appear to be inconsistent with the simulation paper, which states that VPP or HFPP has the highest contrast among the proposed PC methods through defocusing to PP ones. Actually, this kind of inconsistency is found even in the Danev team paper itself, where the VPP PC and defocus PC both taken for the same GPCR species, the contrast of the former appears clearly higher.

As already shown in the discussion about the results shown in Fig. 3 (refer Eqs. 43 to 46), the implication is now clear: the contrast as judged by the human eye is, so to speak, an apparent neglect of high-frequency components. Quantitatively, this means that the missing high-frequency components in the VPP PC linked to low sensitivity appears in the low resolution with the VPP (equivalent to $\pi/2$-ZPP here). This is why defocusing, with its lower apparent contrast but higher sensitivity, is valued in single-particle analysis, where high sensitivity is required, and the results of this paper (especially the result of Eq. 43) are making this point clear. Of course, in the case of low-resolution electron microscopy applications, such as cell observation, VPP or $\pi/2$-ZPP is superior not only in appearance but also in sensitivity, as Eq.46 shows, and in this respect the results in Ref. 3b, which deals with relatively low-resolution electron microscope observation, are self-consistent, where apparent contrast matches sensitivity contrast. Nevertheless, the twin $\varphi$-HPPs method is recommended due to its 1.6-fold sensitivity ratio in the low-resolution range and further sensitivity ratio improvement in the high-frequency range (refer Eqs. 43 to 46).

In this study, the sensitivity is defined based on the CTF cut-off frequency criterion, but as shown implicitly in the Danev team paper [27], the sensitivity criterion based on the information limit seems to be related to the attainable resolution with TEM. Various CTF corrections have been used to extend the frequency region beyond the CTF cut-off frequency and it remains to be seen whether the present conclusions make sense in the case of data analyses with the corrected CTFs [27, 28]. Qualitatively, the conclusions attained here may not change significantly, but quantification of the conclusions will have to await future studies.

In Fig. 4, we find that larger the reduced chromatic aberration $H$ becomes, larger the disagreement between the theory (magenta) and the simulation (blue) becomes. We do not know yet the cause on the discrepancy whether it lies in the theory side or in the simulation side and moreover whether it lies in twin $\varphi$-HPPs or Scherzer defocus. To answer these questions, actual experiments under well-defined parameter values with real TEM equipment must be essential.

Finally, electron dose losses and the charge-induced modulation due to PPs, both of which are the origin of the performance degradation of PP methods, should be revisited. If these drawbacks are eliminated, PP methods can gain more attraction in the TEM community. As the plate in the name of PP implies some kind of matters as the plate material, it looks no way to get rid of electron dose losses and the charging, as the interaction between electrons and matters is just the starting spatial of electron microscopy. However, if we think deeply about the meaning of the phase of the phase-plate, we may need to revisit the wording we have just mentioned. The PP phase reflects the interaction of electrons with electromagnetic potentials involved in the PP materials [29]. To escape the bottleneck, therefore, all you need to do is to have such a phase-plate, which is made of "potentials" but not "materials". This includes, for example, the electrostatic Börsch PP [30], which uses the electrostatic potential of electrodes, the photonic PP using laser, which uses the scalar potential of electromagnetic waves (laser light) [31], and the Aharonov-Bohm (AB) PP [32], which uses vector potentials known as the Aharonov-Bohm effect [33]. Among these, the photonic PP looks most promising as no material exists in the electron beam path. Nevertheless, it may not give the answer to the issue as the PC estimated through simulation was not necessarily high, maybe, due to the complication of the PP phase profile [3b]. On the other hand, the AB HPP [32a, 32c] may have a future as it is a type of HPP able to cover the twin $\varphi$-HPPs and uses only a very thin wire

(less than 1 $\mu m$ in diameter) magnet, in which the electron-matter interaction could be minimized.

## 8. Conclusion

Stimulated by the report that PP PC represented by VPP PC was far inferior to defocus PC in terms of sensitivity [8], we began to investigate ways to increase the sensitivity of PC based on PP methods. As a result, we found a PP scheme superior to the Scherzer defocus if the phase of HPP was reduced from π to around π/2. The reduction of the PP phase from π induces an image of normal-type and differential-type mixed. For the separation of the mixed two images, a two-experiment scheme using a pair of symmetric HPPs were devised and the linear combination of resultant complementary images finally reproduced a high-sensitivity image superior to the Scherzer defocus image, which was termed as the twin $\varphi$-HPPs unified PC image. In the theoretical regime this novel PC scheme resulted in a sensitivity ratio of 1.03-1.73 compared to the Scherzer defocus PC for the experimental setting from low to high-resolution condition and in the simulation regime 1.10-1.38. On the other hand, the sensitivity ratio compared to ZPP was found to increase to 1.58-2.39, in the theoretical regime for the experimental setting from low to high-resolution condition.

**Acknowledgments**

We would like to thank the following individuals for reading this preprint and providing appropriate comments; Haruki Nakamura, Nobuo Tanaka, Radostin Danev, and Fumio Hosokawa,

**Appendix. Definition and Derivation of CTF for twin φ-HPPs unified PC given in Eq. 31**

The CTF is defined first, bearing in mind that the CTF is an entity in spectral space treated in the category of linear imaging.

According to the image intensity modulated with lens aberrations shown in Eq. 27, the PC is expressed by the following equation.

$$\mathcal{F}[\Delta I(r)] = \mathcal{F}[1 - |\mathcal{F}^{-1}[\mathcal{F}[\Psi(r)] P(k) \exp(-i\gamma(k))]|^2] \quad (\Psi(r) = \exp(i\theta(r)) \cong 1 + i\theta(r)),$$
$$= CTF(k)\, \mathcal{F}[\theta(r)]\, A(k). \tag{A1}$$

Here, $A(k)$ corresponds to an aperture function and $\mathcal{F}[\theta(r)]\,A(k)$ to an apertured object in spectral space, of which real space representation is showing up in Eqs. 6 to 12. In the above equation, $CTF(k)$ is a CTF, which reflects the instrumental optical property of $P(k)\exp(-i\gamma(k))$. The CTF is something of a necessary evil. Because, although physically impossible, if the exit wave is directly observed as a complex amplitude, $P(k)\exp(-i\gamma(k))$ itself can be obtained without the bypass of CTF. Complex observation (COBS) is an attempt to make the impossible possible, so to speak, through a two-experiment scheme that is a complexification of TEM experiments to recover $P(k)\exp(-i\gamma(k))$ directly [13].

Next, we attempt to derive the formulation of CTF for twin φ-HPPs unified PC given in Eq.31.
The following four equations are the starting spatial for the derivation.

$$P_L(\mathbf{k}) = A(\mathbf{k}) \left[ (1 + sgn(k_x))/2 + T^{\varphi/\pi} D(k) \exp(i\varphi) (1 - sgn(k_x))/2 \right]. \tag{A2}$$

$$P_R(\mathbf{k}) = A(\mathbf{k}) \left[ (1 - sgn(k_x))/2 + T^{\varphi/\pi} D(k) \exp(i\varphi) (1 + sgn(k_x))/2 \right]. \tag{A3}$$

$$\Delta I_L(\mathbf{r}) = 1 - |\mathcal{F}^{-1}[\mathcal{F}[\Psi(\mathbf{r})] P_L(\mathbf{k}) \exp(-i\gamma(\mathbf{k}))]|^2. \tag{A4}$$

$$\Delta I_R(\mathbf{r}) = 1 - |\mathcal{F}^{-1}[\mathcal{F}[\Psi(\mathbf{r})] P_R(\mathbf{k}) \exp(-i\gamma(\mathbf{k}))]|^2 \quad (\Psi(\mathbf{r}) = \exp(i\theta(\mathbf{r}))). \tag{A5}$$

Eqs. (A1) and (A2) are respectively the same as Eqs. 2 and 8 except for the electron dose losses. $\Delta I_L(\mathbf{r})$ and $\Delta I_R(\mathbf{r})$ respectively corresponds to the left-sided and right-sided $\varphi$-HPP PC. When WPA is applicable both to PPs and objects, two PCs are written as,

$$\Delta I_L(\mathbf{r}) \cong 2\mathcal{I}[|\mathcal{F}^{-1}[P_L(\mathbf{k}) \exp(-i\gamma(\mathbf{k}))]] \otimes \theta(\mathbf{r}), \quad (\because \theta(\mathbf{r}): \text{real}), \tag{A6}$$

$$\Delta I_R(\mathbf{r}) \cong 2\mathcal{I}[|\mathcal{F}^{-1}[P_R(\mathbf{k}) \exp(-i\gamma(\mathbf{k}))]] \otimes \theta(\mathbf{r}), \tag{A7}$$

where $\mathcal{I}[f(\mathbf{r})]$ corresponds to imaginary term extraction for $f(\mathbf{r})$.

As already spatialed out by Koeck for right-sided $\pi/2$-HPP [11], CTFs for $\varphi$-HPP in general have a kind of complex-number forms in spectral space. It implies an image of the normal and differential type mixed in real space and the separation of that mixing is carried out in this paper with a combination of the left-sided and right-sided $\varphi$-HPP and summation and subtraction of resultant PCs. In terms of CTF, thus, they are explicitly shown below as,

$$1/2 \mathcal{F}[\Delta I_L(\mathbf{r}) + \Delta I_R(\mathbf{r})] = A(\mathbf{k})[-\sin\gamma(\mathbf{k}) + T^{\varphi/\pi} D(k) \sin(\varphi - \gamma(\mathbf{k}))] \mathcal{F}[\theta(\mathbf{r})], \tag{A8}$$

$$1/2 \mathcal{F}[\Delta I_L(\mathbf{r}) - \Delta I_R(\mathbf{r})] = A(\mathbf{k})[\cos\gamma(\mathbf{k}) - T^{\varphi/\pi} D(k) \cos(\varphi - \gamma(\mathbf{k}))](-i\, sgn(k_x))\mathcal{F}[\theta(\mathbf{r})], \tag{A9}$$

where the property that $A(\mathbf{k})$, $D(k)$ and $\gamma(\mathbf{k})$ are even functions is used. Note that $1/2\mathcal{F}[\Delta I_L(\mathbf{r}) + \Delta I_R(\mathbf{r})]$ is real and $1/2\mathcal{F}[\Delta I_L(\mathbf{r}) - \Delta I_R(\mathbf{r})]$ imaginary. The beauty of twin φ-HPPs unified method is that it does not force the complex-number CTF to be real with an inverse filtering [11], but rather separates the CTF into pure real and pure imaginary terms, as seen above, in response to the real-imaginary separation of the phase disturbance function $\exp(-i\gamma(\mathbf{k}))$. This is precisely the basic principle of COBS [13], which underpins the argument developed in Discussion to bring the twin φ-HPPs unified method into the category of COBS.

Hilbert transform can take things a step further and eliminate the factor ($-i\,sgn(k_x)$) found in Eq. A8. Finally, we have following,

$$1/2\mathcal{F}[\Delta I_L(r) + \Delta I_R(r)] + 1/2\mathcal{F}[H[\Delta I_L(r) - \Delta I_R(r)]]$$

$$= A(\mathbf{k})[\cos\gamma(\mathbf{k}) - \sin\gamma(\mathbf{k}) + T^{\varphi/\pi}D(\mathbf{k})(\sin(\varphi - \gamma(\mathbf{k})) - \cos(\varphi - \gamma(\mathbf{k}))]\mathcal{F}[\theta(r)],$$

$$= [\cos\gamma(\mathbf{k}) - \sin\gamma(\mathbf{k}) + T^{\varphi/\pi}D(\mathbf{k})(\sin(\varphi - \gamma(\mathbf{k})) - \cos(\varphi - \gamma(\mathbf{k}))]A(\mathbf{k})\mathcal{F}[\theta(r)],$$

$$= \sqrt{2}(\cos(-\gamma(\mathbf{k}) - \pi/4) + T^{\varphi/\pi}D(k)\sin(-\gamma(\mathbf{k}) + \varphi - \pi/4))A(\mathbf{k})\mathcal{F}[\theta(r)], \quad (A10)$$

$$CTF(\mathbf{k}) = \sqrt{2}(\cos(-\gamma(\mathbf{k}) - \pi/4) + T^{\varphi/\pi}D(k)\sin(-\gamma(\mathbf{k}) + \varphi - \pi/4)) \quad (A11)$$

where $H[\ ]$ indicates a sp corresponding to the Hilbert transform. Though the path to the final image of twin φ-HPPs unified PC is visually shown in Figs. 5 and 6, its theoretical essence is information compressed into Eqs. A10 and A11.

**Figure Captions**

Fig. 1, Variations of the Hilbert phase-plate (HPP).
a, Original HPP with PP phase π (termed as π-HPP)
b, HPP with PP phase $\varphi$ ($\varphi$ is smaller than π) (termed as $\varphi$-HPP)
c, Right-sided $\varphi$-HPP with the PP position set to the right of the optical axis (termed as right sided $\varphi$-HPP)
d, HPP, which simultaneously holds symmetrical two $\varphi$-HPPs (left-sided and right-sided $\varphi$-HPP) (termed as twin $\varphi$-HPPs).

Fig. 2, 3-function product CTF (3FP-CTF) (a) and 4-function product CTF (4FP-CTF) (b) for the case of reduced chromatic aberration $H$=5 nm.

Fig. 3, Sensitivity ratios between two PC methods in the theoretical and simulation regime. Twin $\varphi$-HPPs PC vs Scherzer defocus PC (theoretical): magenta diamond, $\varphi$-ZPP PC vs Scherzer defocus PC (theoretical): blue triangle, twin $\varphi$-HPPs vs Scherzer defocus (simulated): red circle.
a, for $H$=5 nm; b, for $H$=20 nm; c, for $H$=40 nm; d, for $H$=100 nm.

Fig. 4, Comparison of two 4FP-CTFs to search for the optimum parameter $H$ involved in the chromatic aberration envelope function $K_C(k)$. The magenta line shows the 4FP-CTF without DQE-related electron dose loss, the blue line shows the 4FP-CTF with DQE-related electron dose loss. The parameter $H$ for the 4FP-CTF represented by the magenta line (namely 4FP-CTF without DQE) was deliberately increased to adjust integrations of two 4FP-CTFs without DQE and with DQE being equivalent.  a, for $H$=5 nm; b, for $H$=20 nm; c, for $H$=40 nm; d, for $H$=100 nm.

Fig. 5, Simulated images of a single gold atom and an illustration of the flowchart of algorithm to find the twin $\varphi$-HPPs unified image in the case of the highest sensitivity ($\varphi$=1.29 rad) for $H$=5 nm ($\varphi$=1.29 rad). Top row: images corresponding to the left-sided and right-sided $\varphi$-HPP, second row: the sum and subtract image, third row: Hilbert transformed image of the subtract image, bottom row: left; Scherzer defocus image, center; twin $\varphi$-HPPs unified image, right; π/2-ZPP image.

Fig. 6, Simulated images of a single gold atom and an illustration of the flowchart of algorithm to find the twin $\varphi$-HPPs unified image in the case of the highest sensitivity ($\varphi$=1.80 rad) for $H$=100 nm. Top row: images corresponding to the left-sided and right-sided $\varphi$-HPP, second row: the sum and subtract image, third row: Hilbert transformed image of the subtract image, bottom row: left; Scherzer defocus image, center; twin $\varphi$-HPPs unified image, right; π/2-ZPP image.

Fig. 1

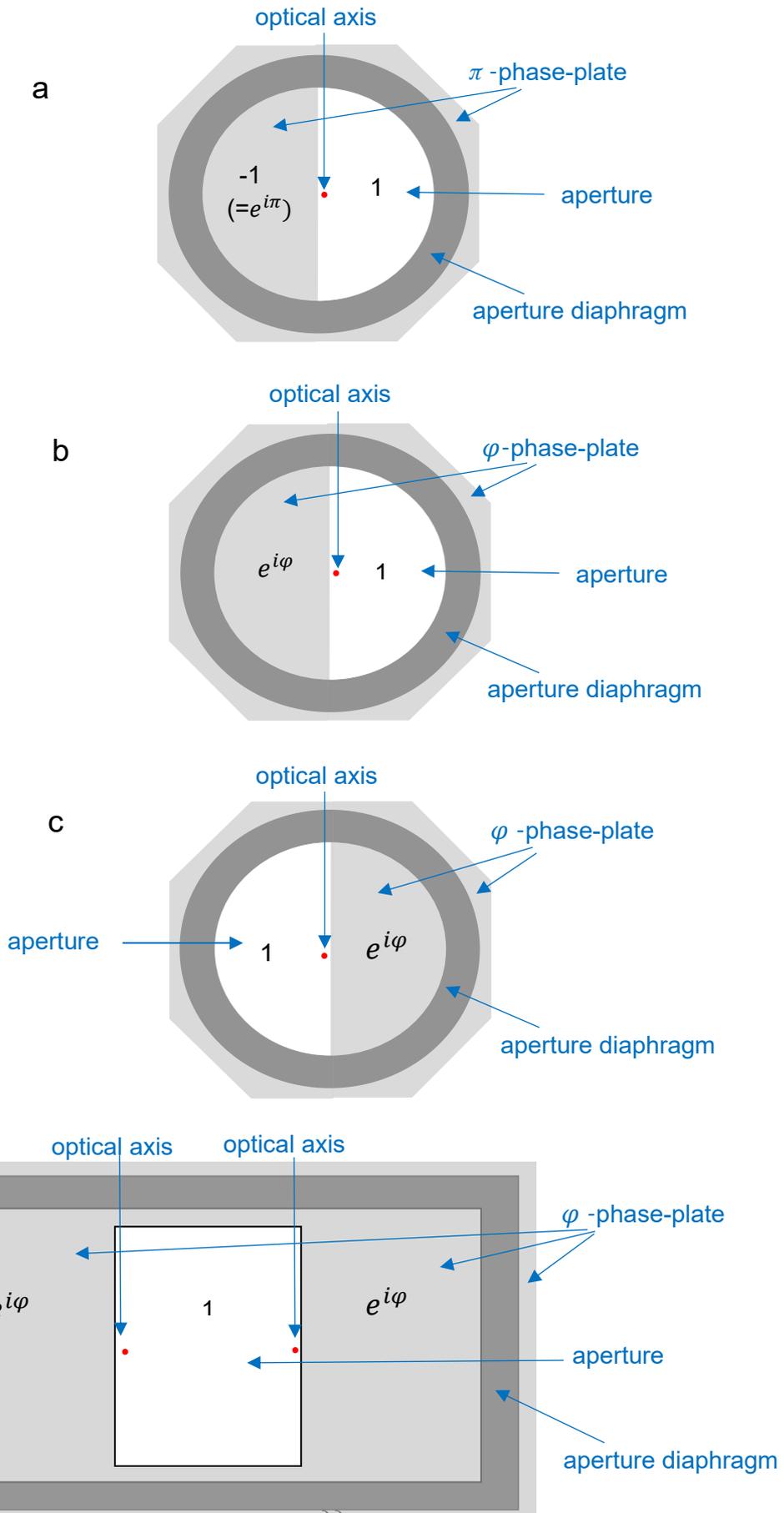

Fig. 2

a
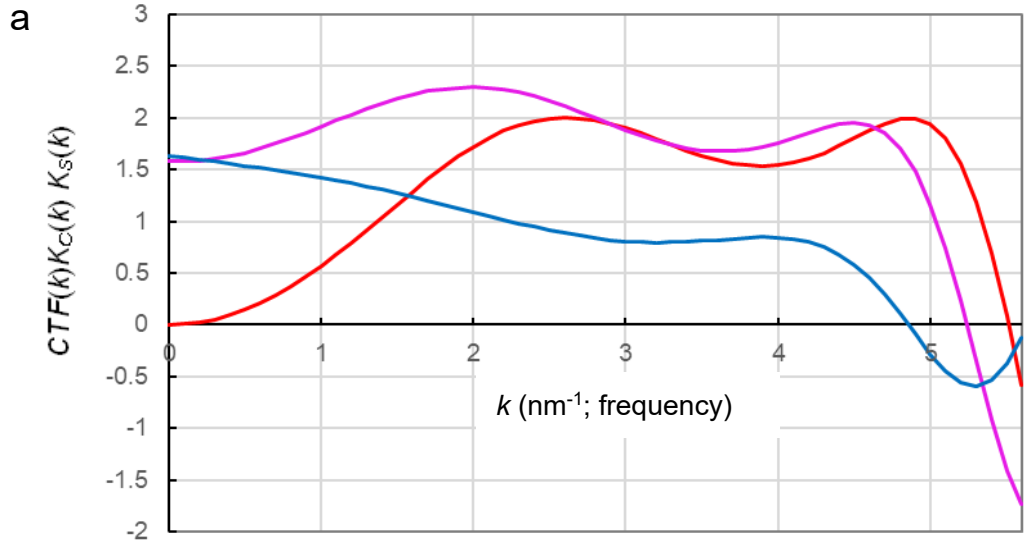

b
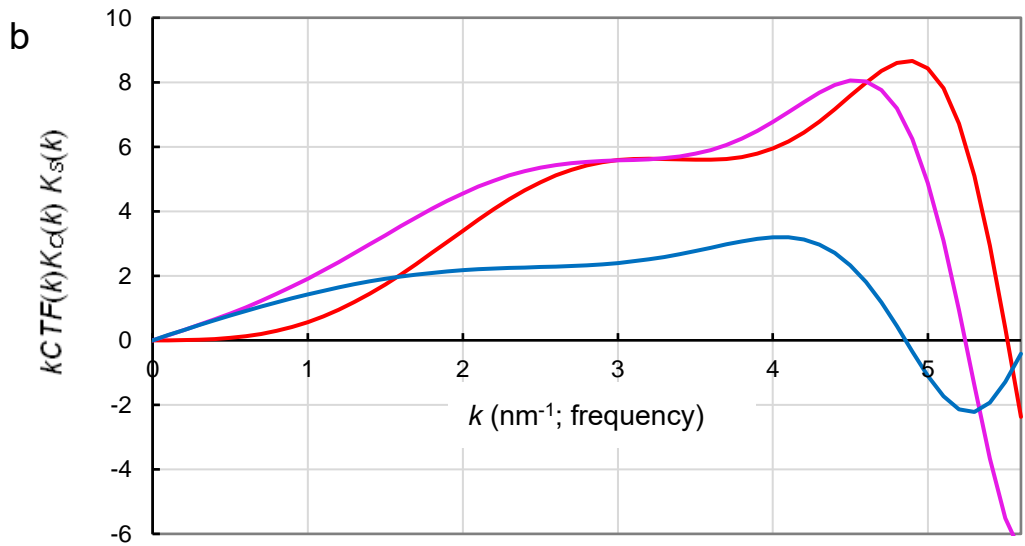

Fig. 3

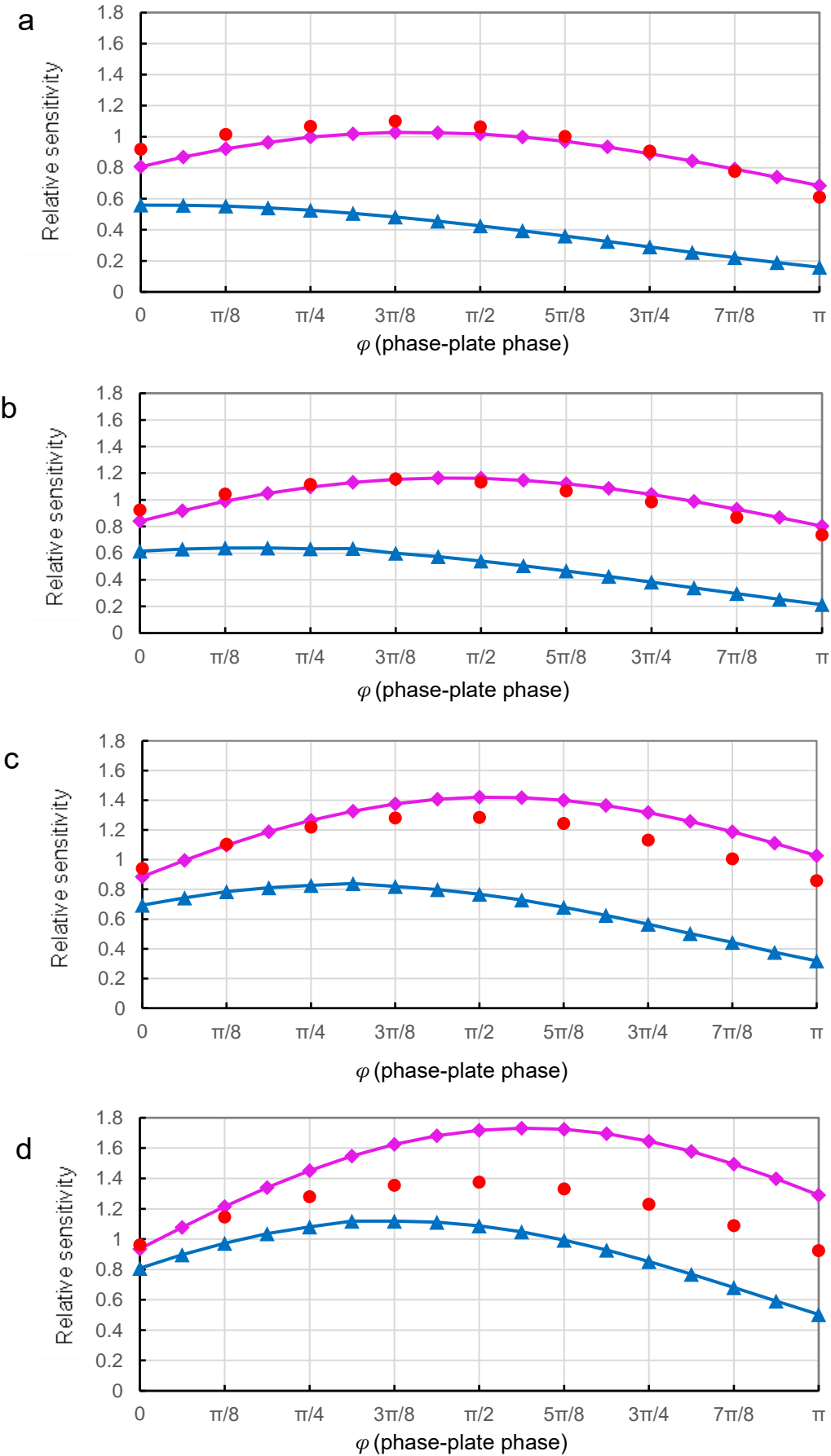

Fig. 4

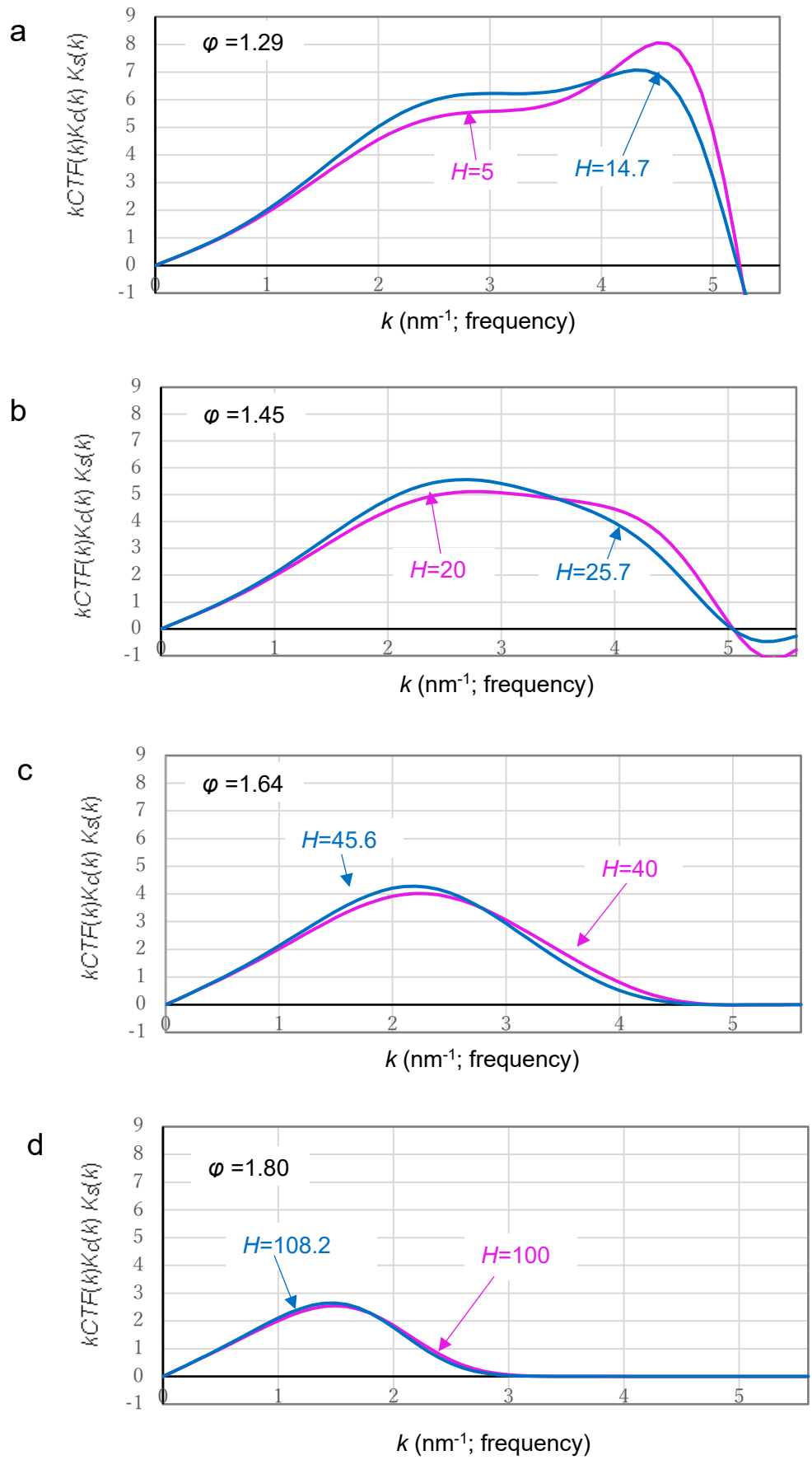

Fig. 5

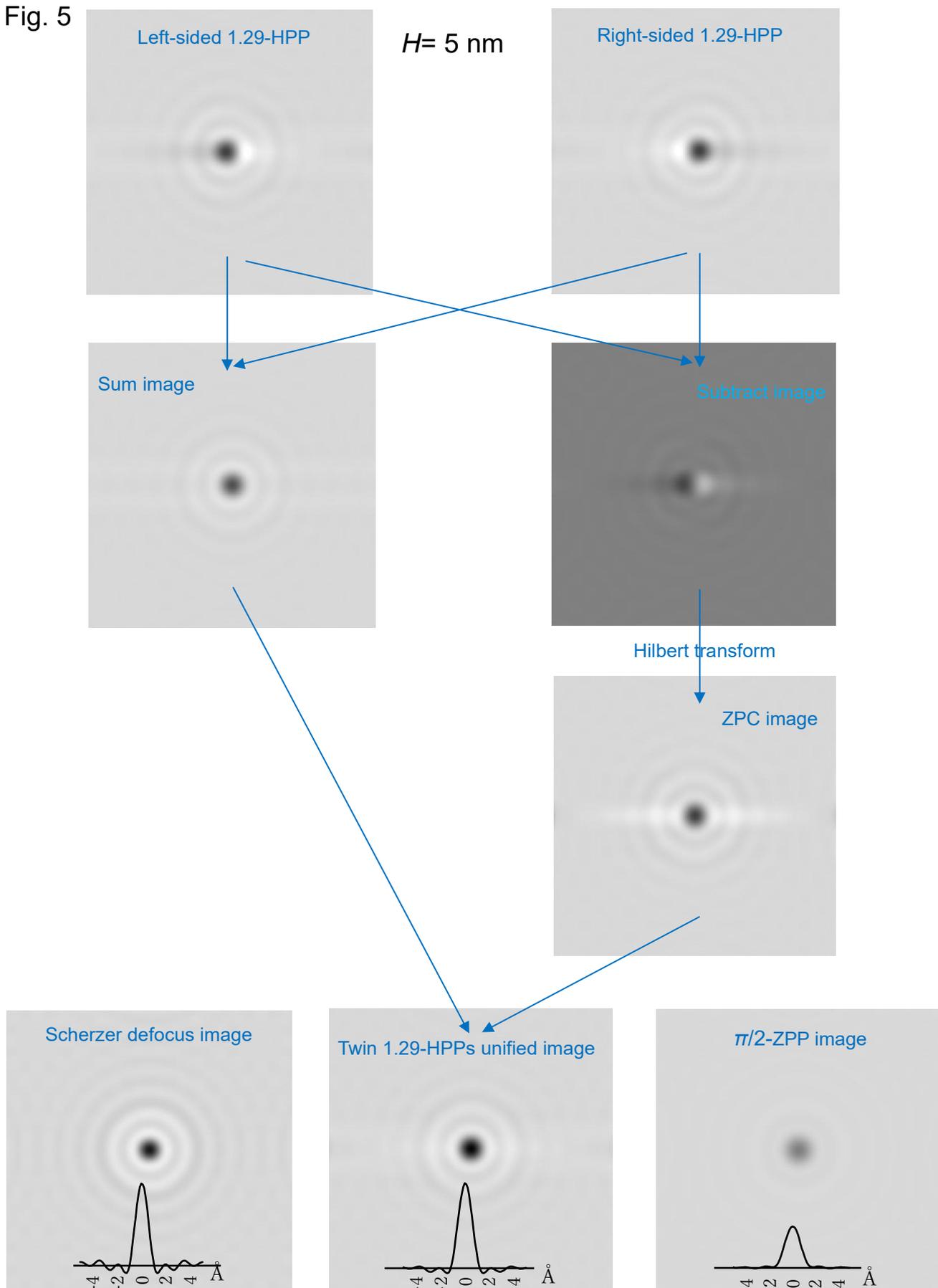

Fig. 6

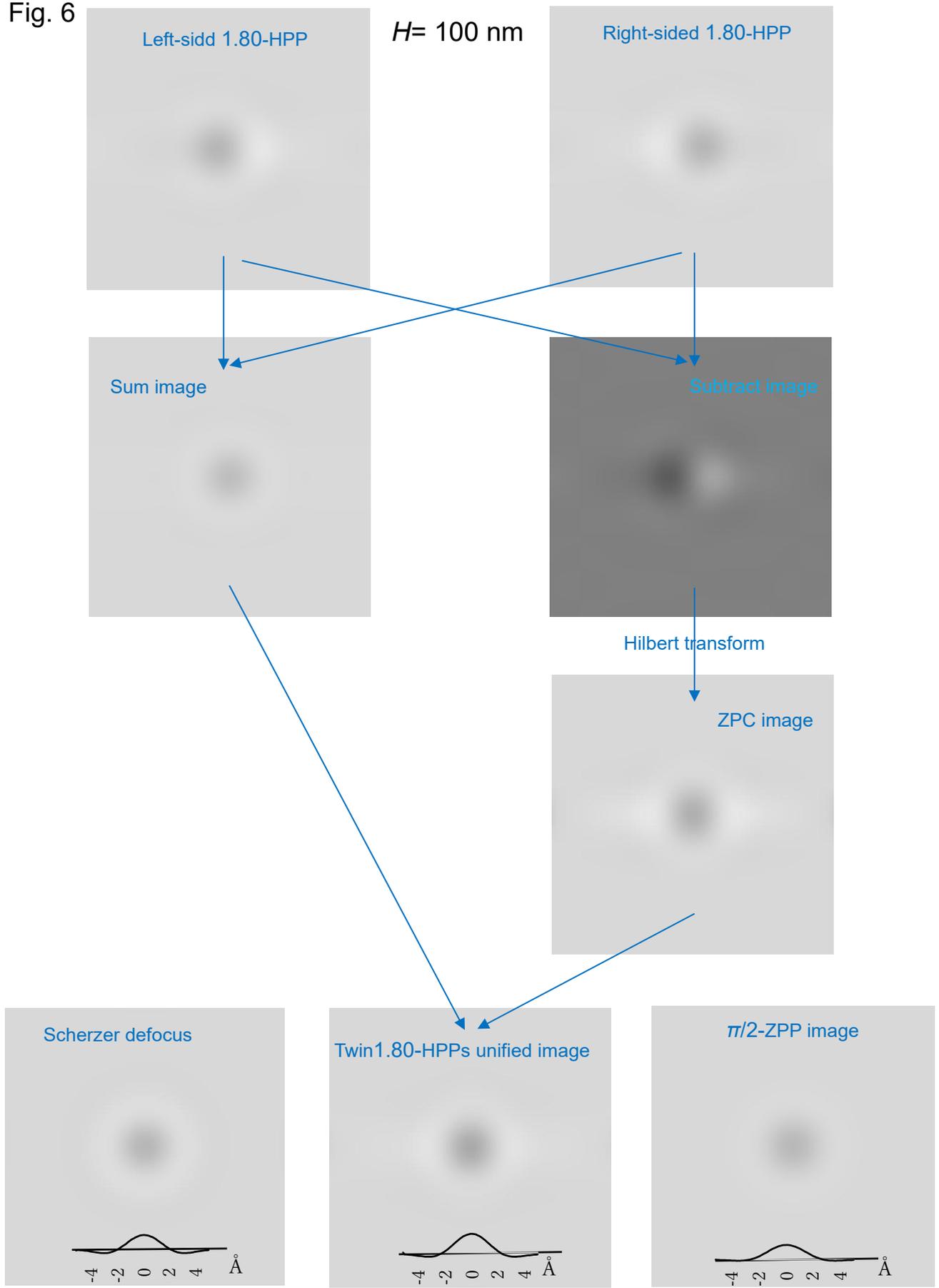